   \documentclass[twocolumn,showpacs,preprintnumbers,amsmath,amssymb]{revtex4}
   \usepackage{graphicx} 
   \usepackage{bm}
   \usepackage{dcolumn}
   \usepackage{epstopdf}
   \usepackage{natbib}

\begin{document}
   	
   	\title{Spectral properties of excitons in the bilayer graphene}
   	
   	\author{V. Apinyan\footnote{Corresponding author. Tel.:  +48 71 3954 284; E-mail address: v.apinyan@int.pan.wroc.pl.}, T. K. Kope\'{c}}
   	\affiliation{Institute for Low Temperature and Structure Research, Polish Academy of Sciences\\
   		PO. Box 1410, 50-950 Wroc\l{}aw 2, Poland \\}
   	
   	\date{\today}

\begin{abstract}
%
In this paper, we consider the spectral properties of the bilayer graphene with the local excitonic pairing interaction between the electrons and holes. We consider the generalized Hubbard model, which includes both intralayer and interlayer Coulomb interaction parameters. The solution of the excitonic gap parameter is used to calculate the electronic band structure, single-particle spectral functions, the hybridization gap, and the excitonic coherence length in the bilayer graphene. We show that the local interlayer Coulomb interaction is responsible for the semimetal-semiconductor transition in the double layer system, and we calculate the hybridization gap in the band structure above the critical interaction value. The formation of the excitonic band gap is reported as the threshold process and the momentum distribution functions have been calculated numerically. We show that in the weak coupling limit the system is governed by the Bardeen-Cooper-Schrieffer (BCS)-like pairing state. Contrary, in the strong coupling limit the excitonic condensate states appear in the semiconducting phase, by forming the Dirac's pockets in the reciprocal space.      
   	\end{abstract}

   	\maketitle

 \renewcommand\thesection{\arabic{section}}
   	
\section{\label{sec:Section_1} Introduction}
%

The electronic band gap of semiconductors and insulators largely determines their optical, transport properties and governs the operation of semiconductor based devices such as p-n junctions, transistors, photodiodes and lasers \cite{cite_1}. Opening up a band gap in the bilayer graphene (BLG), by applying the external electric field and finding a suitable substrate are two challenges for constituting the modern nano-electronic equipment \cite{cite_2, cite_3}. The imposition of external electrical field can tune the bilayer graphene from the semimetal to the semiconducting state \cite{cite_2}.
On the other hand, the possibility of formation of the excitonic insulator state and the excitonic condensation in the bilayer graphene structures remains controversial in the modern solid state physics \cite{cite_4, cite_5,cite_6,cite_7,cite_8,cite_9,cite_10, cite_11,cite_12,cite_13,cite_14}. In difference with the quasi two-dimensional (2D) semiconducting systems, where those two states have been observed experimentally and well discussed theoretically \cite{cite_15, cite_16,cite_17,cite_18,cite_19,cite_20,cite_21, cite_22,cite_23,cite_24,cite_25, cite_26,cite_27}, the formation of the excitonic condensate states in the BLG system, from the original electron-hole pairing states, is much more obscure because of the complicated nature of the single-particle correlations in these systems \cite{cite_6,cite_8,cite_10,cite_13,cite_14}. The weak correlation diagrammatic mechanism, discussed in the Refs.\onlinecite{cite_13,cite_14}, is restricted only to the closed loop expansion in the diagrammatic series, and in this case, only the density fluctuation effects could affect the formation of the excitonic condensate states. Meanwhile, it has been shown \cite{cite_28, cite_29, cite_30} that even the undoped graphene can provide a variety of electron-hole type pairing chiral symmetry breaking orders especially for the strong Coulomb coupling case, which renders the treatments in Refs.\onlinecite{cite_13,cite_14} to be nontrivial. As the monolayer graphene, bilayer graphene has a semimetallic band structure with the zero bandgap, which is unsuitable for many electronic device applications, as routinely done with semiconducting devices. In $2007$, Allan H. MacDonald and his colleagues have predicted that the electric displacement field, applied to the two layers of the BLG could introduce a tunable bandgap in the electronic band structure of the BLG \cite{cite_31}, which has been proved later on, experimentally \cite{cite_32,cite_33,cite_34}.
   	
In this paper, we show the band gap formation in the bilayer graphene's band structure, mediated by the local interlayer Coulomb interaction parameter and without the external electric field. Particularly, we show how the interlayer Coulomb interaction tunes the BLG from the semimetallic state into the semiconducting one (or a possible insulator state, for large interaction limit), for a fixed value of the interlayer hopping amplitude. We will show that the formation of the band gap in the BLG is a threshold process in our case. We calculate the excitonic hybridization gap for different values of the interaction parameter. Furthermore, by using the exact solutions for the excitonic gap parameter and chemical potential, we calculate the single-particle spectral functions and the momentum distribution functions in the BLG for different values of the local interlayer interaction parameter. We will show that the system is governed by the weak-coupling Bardeen-Cooper-Schrieffer (BCS)-like pairing states for the small and intermediate values of the interlayer coupling parameter, and the behavior of the excitonic coherence length agrees well with the BCS type relation $\xi_c\sim\bar{\mu}/({k_{\cal{F}}}\Delta)$, with $\Delta$, being the excitonic pairing gap parameter. Contrary, in strong coupling limit the coherence length becomes proportional to $\Delta$ ($\xi_c\sim\Delta)$, and the system is in the excitonic condensate regime. 
   	
In the Section \ref{sec:Section_2}, we introduce the bilayer Hubbard model for the BLG system. In the Section \ref{sec:Section_3}, we give the action of the bilayer graphene system and we obtain the coupled self-consistent equations for the excitonic gap parameter and chemical potential. The excitonic dispersion relations are also given there. Next, in the Section \ref{sec:Section_4}, we discuss the single-particle spectral properties and momentum distribution functions for different interlayer interaction limits and we calculate the interlayer excitonic coherence length in the bilayer graphene. In the Section \ref{sec:Section_5}, we give a conclusion to our paper.
   	
   	\section{\label{sec:Section_2} The method}
\begin{figure}
   		\begin{center}
   			\includegraphics[width=220px,height=140px]{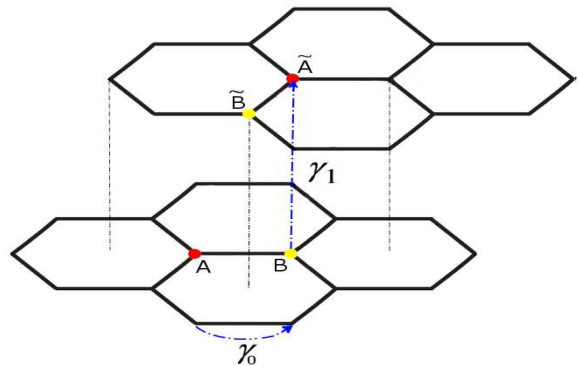}
   			\caption{\label{fig:Fig_1}(Color online) 
   				The bilayer graphene structure. Two different sublattice site positions are shown ($A$,$B$, and $\tilde{A}$, $\tilde{B}$) in two different layers of the bilayer graphene structure.}
   		\end{center}
   	\end{figure}

   	%
   	The BLG system, considered here, is composed of two coupled honeycomb layers with sublattices $A$, $B$ and $\tilde{A}$, $\tilde{B}$, in the bottom and top layers respectively (see in Fig.~\ref{fig:Fig_1}). In the $z$-direction the layers are arranged according to Bernal Stacking (BS) order \cite{cite_35}, i.e. the atoms on the sites $\tilde{A}$ in the top layer lie just above the atoms on the sites $B$ in the bottom layer graphene, and each layer is composed of two interpenetrating triangular lattices.
   	For a simple treatment at equilibrium, we initially suppose the balanced BLG structure, i.e., with the equal chemical potentials in the both layers. When switching the local Coulomb potential $W$ between the layers, we keep the charge neutrality equilibrium through the BLG, by imposing the half-filling condition in each layer of the BLG. Next, we will pass to the Grassmann representation for the fermionic variables, and we write the partition function of the system, in the imaginary time fermion path integral formalism \cite{cite_36}.
   	For this, we introduce the imaginary-time variables $\tau$, at each lattice site ${\bf{r}}$. The time variables $\tau$ vary in the interval $(0,\beta)$, where $\beta=1/T$ with $T$ being the temperature. Then, the Hamiltonian of the bilayer graphene, with the local interlayer interaction, has the following form
   	\begin{eqnarray}
   	H&=&-\gamma_0\sum_{\left\langle {\bf{r}}{\bf{r}}'\right\rangle}\sum_{\sigma}\left({a}^{\dag}_{\sigma}({\bf{r}})b_{\sigma}({\bf{r}}')+h.c.\right)
   	\nonumber\\
   	&-&\gamma_0\sum_{\left\langle {\bf{r}}{\bf{r}}'\right\rangle}\sum_{\sigma}\left({\tilde{b}}^{\dag}_{\sigma}({\bf{r}})\tilde{b}_{\sigma}({\bf{r}}')+h.c.\right)
   	\nonumber\\
   	&-&\gamma_1\sum_{{\bf{r}}\sigma}\left({{b}}^{\dag}_{\sigma}({\bf{r}})\tilde{a}_{\sigma}({\bf{r}})+h.c.\right)-\sum_{{\bf{r}}\sigma}\sum_{\ell=1,2}\mu_{\ell}n_{\ell\sigma}({\bf{r}})
   	\nonumber\\
   	&+&U\sum_{{\bf{r}}}\sum_{\ell\eta}\left[\left(n_{\ell\eta\uparrow}-1/2\right)\left(n_{\ell\eta\downarrow}-1/2\right)-1/4\right]
   	\nonumber\\
   	&+&W\sum_{{\bf{r}}\sigma\sigma'}\left[\left(n_{1b\sigma}({\bf{r}})-1/2\right)\left(n_{2\tilde{a}\sigma'}({\bf{r}})-1/2\right)-1/4\right].
   	\nonumber\\
   	\label{Equation_1}
   	\end{eqnarray}
Here, we have used the graphite nomenclature notations \cite{cite_35} for the hopping amplitudes $\gamma_{0}$ and $\gamma_{1}$. Namely, the parameter $\gamma_{0}$ is the intraplane hopping amplitude, and $\gamma_{1}$ is the interlayer hopping amplitude in the BLG (see also in Fig.~\ref{fig:Fig_1}). The summation $\left\langle {\bf{r}}{\bf{r}}' \right\rangle$, in the first term, in Eq.(\ref{Equation_1}), denotes the sum over the nearest neighbors lattice sites in the separated honeycomb layers in the bilayer graphene structure. We keep the small letters $a,b$ and $\tilde{a}, \tilde{b}$ for the electron operators on the lattice sites $A,B$ and $\tilde{A},\tilde{B}$ respectively (see in Fig.~\ref{fig:Fig_1}).
The index $\ell=1,2$ denotes the number of single layers in the BLG. Particularly, we use $\ell=1$ for the bottom layer, and $\ell=2$ for the top layer. Furthermore, we have $\eta=a, b$ for $\ell=1$, and $\eta=\tilde{a}, \tilde{b}$, for $\ell=2$. The symbol $\sigma$ denotes the spin variables with two possible directions ($\sigma= \uparrow, \downarrow$). 
Next, $n_{\ell\sigma}({\bf{r}})$ is the total electron density operator for the layer $\ell$ with a given spin direction $\sigma$
\begin{eqnarray}
n_{\ell\sigma}({\bf{r}})=\sum_{\eta}n_{\ell\eta\sigma}({\bf{r}}),
\label{Equation_2}
\end{eqnarray}
and $n_{\ell\eta\sigma}({\bf{r}})={\eta}^{\dag}_{\ell\sigma}({\bf{r}})\eta_{\ell\sigma}({\bf{r}})$ is the electron density operator for the $\eta$-type fermions with the spin $\sigma$.
We consider the BLG structure with pure electronic layers without initial dopping in the system. The condition of half-filling in each layer reads as $\left\langle n_{\ell} \right\rangle=1$, for $\ell=1,2$, where $n_{\ell}$ is the total electron density operator for the layer $\ell$ summed over the spin index $\sigma$: $n_{\ell}({\bf{r}})=\sum_{\sigma}n_{\ell\sigma}({\bf{r}})$. Furthermore, $U$, in the Hubbard term in Eq.(\ref{Equation_1}), parametrizes the intralayer Coulomb interaction. 
The parameter $W$, in the last term in Eq.(\ref{Equation_1}), describes the local interlayer Coulomb repulsion between the electrons located on the $B$ and $\tilde{A}$ type of sites in the different layers of the BLG. 	We will put $\gamma_{0}=1$, as the unit of energy in the system, and we set $k_{B}=1$, $\hbar=1$ throughout the paper.
   	%
   	%
   	\section{\label{sec:Section_3} The Green's function matrix}
   	%
The pairing between the electron and holes results in a gap in the excitation energy spectrum of the system. The pairing gap parameter between the particles with the same spin orientations is $\Delta_{\sigma\sigma'}=\Delta_{\sigma\sigma}\delta_{\sigma\sigma'}$, and we can also assume that the pairing gap is real $\Delta_{\sigma\sigma}={\Delta}^{\dag}_{\sigma\sigma}\equiv\Delta$. Then the excitonic pairing gap parameter will be defined as 
   	\begin{eqnarray}
   	\Delta_{\sigma\sigma}=W\left\langle {b}^{\dag}_{\sigma}({\bf{r}}\tau)\tilde{a}_{\sigma}({\bf{r}}\tau)\right\rangle.
   	\label{Equation_3}
   	\end{eqnarray}
In order to find the momentum dependence of the excitonic gap parameter, we will pass to the Fourier space representation, given by the transformation
	\begin{eqnarray}
	\eta_{\sigma}({\bf{r}},\tau)=\frac{1}{\beta{N}}\sum_{{\bf{k}}\nu_{n}}\eta_{\sigma{\bf{k}}}(\nu_{n})e^{i\left({\bf{k}}{\bf{r}}-\nu_{n}\tau\right)},
	\nonumber\\
	\end{eqnarray}
	where $N$ is the total number of sites on the $\eta$-type sublattice, in the layer $\ell$, and we write the fermionic action of the bilayer graphene system in the form 
   	
   	\begin{eqnarray} 
   	{\cal{S}}\left[{\psi}^{\dag},\psi\right]=\frac{1}{\beta{{ N}}}\sum_{{\bf{k}}\nu_{n}}\sum_{\sigma}{\psi}^{\dag}_{\sigma{\bf{k}}}(\nu_{n}){\hat{\cal{G}}}^{-1}_{\sigma{\bf{k}}}(\nu_{n}){\psi}_{\sigma{\bf{k}}}(\nu_{n}).
   	\label{Equation_4}
   	\end{eqnarray}
   	Here, $\nu_{n}=\pi\left(2n+1\right)/\beta$ with $n=0,\pm1,\pm2,\dots$, are the fermionic Matsubara frequencies \cite{cite_37}. The four component Dirac spinors $\psi_{\sigma{\bf{k}}}(\nu_{n})$, in Eq.(\ref{Equation_4}), have been introduced at each discrete state ${\bf{k}}$ in the reciprocal space and for each spin direction $\sigma=\uparrow, \downarrow$. Being the generalized Weyl spinors, they are defined as
   	
   	\begin{eqnarray} {\psi}_{\sigma{\bf{k}}}(\nu_{n})=\left[a_{\sigma{\bf{k}}}(\nu_{n}),b_{\sigma{\bf{k}}}(\nu_{n}),\tilde{a}_{\sigma{\bf{k}}}(\nu_{n}),\tilde{b}_{\sigma{\bf{k}}}(\nu_{n})\right]^{T}.
   	\label{Equation_5}
   	\end{eqnarray}
   	
   	The matrix ${\hat{{\cal{G}}}}^{-1}_{\sigma{\bf{k}}}(\nu_{n})$, in Eq.(\ref{Equation_4}), is the inverse Green's function matrix, of size $4\times4$. It is defined as
   	\begin{eqnarray}
   	\footnotesize
   	\arraycolsep=0pt
   	\medmuskip = 0mu
   	{\hat{{\cal{G}}}}^{-1}_{\sigma{\bf{k}}}\left(\nu_{n}\right)=\left(
   	\begin{array}{ccccrrrr}
   	E_{1}(\nu_{n}) & -\tilde{\gamma}_{1{\bf{k}}} & 0 & 0\\
   	-\tilde{\gamma}^{\ast}_{1{\bf{k}}} &E_{2}(\nu_{n})  & -\gamma_{1}-{\Delta}^{\dag}_{\sigma} & 0 \\
   	0 & -\gamma_{1}-{\Delta}_{\sigma} & E_{2}(\nu_{n}) & -\tilde{\gamma}_{2{\bf{k}}} \\
   	0 & 0 & -\tilde{\gamma}^{\ast}_{2{\bf{k}}} & E_{1}(\nu_{n}) 
   	\end{array}
   	\right).
   	\label{Equation_6}
   	\end{eqnarray}
   	Indeed, the structure of the matrix does not changes when inverting the spin direction, i.e., ${\hat{{\cal{G}}}}^{-1}_{\downarrow{\bf{k}}}\left(\nu_{n}\right)\equiv {\hat{{\cal{G}}}}^{-1}_{\uparrow{\bf{k}}}\left(\nu_{n}\right)$.
   	The diagonal elements of the matrix, in Eq.(\ref{Equation_6}), are the single-particle quasienergies 
   	\begin{eqnarray}
   	E_{\ell}(\nu_{n})=-i\nu_{n}-\mu^{\rm eff}_{\ell},
   	\label{Equation_7}
   	\end{eqnarray}
   	where the effectve chemical potentials $\mu^{\rm eff}_{\ell}$ with $\ell=1,2$ have been introduced as $\mu^{\rm eff}_{1}=\mu+U/4$ and $\mu^{\rm eff}_{2}=\mu+U/4+W$. The parameters $\tilde{\gamma}_{\ell{\bf{k}}}$, in Eq.(\ref{Equation_6}), are the renormalized (nearest neighbors) intralayer hopping amplitudes and $\tilde{\gamma}_{\ell{\bf{k}}}=z\gamma_{\ell{\bf{k}}}\gamma_0$, where the ${\bf{k}}$-dependent parameters $\gamma_{\ell{\bf{k}}}$ are the usual tight-binding energy dispersions in the BLG. Namely, we have 
   	\begin{eqnarray}
   	\gamma_{\ell{\bf{k}}}=\frac{1}{z}\sum_{\vec{{\bf{\delta}}}_{\ell}}e^{-i{{\bf{k}}\vec{{\bf{\delta}}}_{\ell}}}.
   	\label{Equation_8}
   	\end{eqnarray}
   	The parameter $z$, is the number of the nearest neighbors lattice sites on the honeycomb lattice. The components of the nearest-neighbors vectors ${\vec{\bf{\delta}}}_{\ell}$, for the bottom layer 1, are given by ${\vec{\bf{\delta}}}^{(1)}_{1}=\left(d/2,d\sqrt{3}/2\right)$, ${\vec{\bf{\delta}}}^{(2)}_{1}=\left(d/2,-d\sqrt{3}/2\right)$ and ${\vec{\bf{\delta}}}^{(3)}_{1}=\left(-d,0\right)$. For the layer 2, we have obviously ${\vec{\bf{\delta}}}^{(1)}_{2}=\left(d,0\right)$, ${\vec{\bf{\delta}}}^{(2)}_{2}=\left(-d/2,-d\sqrt{3}/2\right)$, and ${\vec{\bf{\delta}}}^{(3)}_{2}=\left(-d/2,d\sqrt{3}/2\right)$. Then, for the function $\gamma_{1{\bf{k}}}$, we get
   	\begin{align}
   	\gamma_{1{\bf{k}}}=\frac{1}{3}\left[e^{-ik_{x}d}+2e^{i\frac{k_{x}d}{2}}\cos\left(\small{\frac{\sqrt{3}}{2}k_{y}d}\right)\right],
   	\label{Equation_9}
   	\end{align}
       where $d$ is the carbon-carbon interatomic distance. It is not difficult to realize that $\gamma_{2{\bf{k}}}=\gamma^{\ast}_{1{\bf{k}}}\equiv\gamma^{\ast}_{{\bf{k}}}$, and therefore we have $\tilde{\gamma}_{2{\bf{k}}}=\tilde{\gamma}^{\ast}_{1{\bf{k}}}\equiv\tilde{\gamma}^{\ast}_{{\bf{k}}}$.
       The partition function of the bilayer graphene system will read as
   	\begin{eqnarray}
   	Z=\int\left[{\cal{D}}{\psi}^{\dag}{\cal{D}}\psi\right]e^{-{\cal{S}}\left[{\psi}^{\dag},\psi\right]},	
   	\label{Equation_10}
   	\end{eqnarray}
   	where the action of the system is given in Eq.(\ref{Equation_4}) above. 
   	The form of the Green's function matrix, given in Eq.(\ref{Equation_6}), will be used in the next Section, in order to derive the self-consistent equations, which determine the excitonic gap parameter $\Delta$ and the effective bare chemical potential $\bar{\mu}$ in the interacting BLG system. Particularly, this last one plays an important role in the BLG theory and redefines the charge neutrality point (CNP) \cite{cite_38,cite_39, cite_40} in the context of the exciton formation in the interacting bilayer graphene. Quite interesting experimental results on that subject are given recently in Refs.\onlinecite{cite_38,cite_39, cite_40}.
   	%
   	\subsection{\label{sec:Section_3_1} The excitonic dispersion relations}
   	%
   	We will perform the Hubbard-Stratanovich transformation of the partition function in Eq.(\ref{Equation_10}). In the Dirac's spinor notations, the partition function will be transformed as follows
   	\begin{eqnarray}
   	Z=\int\left[{\cal{D}}{\psi}^{\dag}{\cal{D}}\psi\right]e^{-\frac{1}{\beta{{ N}}}\sum_{{\bf{k}}\nu_{n}}\sum_{\sigma}{\psi}^{\dag}_{\sigma{\bf{k}}}(\nu_{n}){\hat{\cal{G}}}^{-1}_{\sigma{\bf{k}}}(\nu_{n}){\psi}_{\sigma{\bf{k}}}(\nu_{n})}\times
   	\nonumber\\
   	\times 
   	e^{\frac{1}{\beta{ N}}\sum_{{\bf{k}}\nu_{n}}\sum_{\sigma}\left[\frac{1}{2}{J}^{\dag}_{{\bf{k}}\sigma}(\nu_{n})\psi_{{\bf{k}}\sigma}(\nu_{n})+\frac{1}{2}{\psi}^{\dag}_{{\bf{k}}\sigma}(\nu_{n}){J}_{{\bf{k}}\sigma}(\nu_{n})\right]}\approx
   	\nonumber\\
   	\approx e^{\frac{\beta{N}}{4}\sum_{{\bf{k}}\nu_{n}}\sum_{\sigma}{J}^{\dag}_{{\bf{k}}}(\nu_{n}){\hat{\cal{G}}}_{\sigma{\bf{k}}}(\nu_{n}){J}_{{\bf{k}}\sigma}(\nu_{n})},
   	\nonumber\\
   	\label{Equation_11}
   	\end{eqnarray}
   	where we have introduced the auxiliary fermionic source field vectors ${J}_{{\bf{k}}\sigma}(\nu_{n})$, which are also Dirac's spinors as the $\psi$-fields, defined in the Section \ref{sec:Section_3}
   	\begin{eqnarray} J_{\sigma{\bf{k}}}(\nu_{n})=\left[j_{a\sigma{\bf{k}}}(\nu_{n}),j_{b\sigma{\bf{k}}}(\nu_{n}),j_{\tilde{a}\sigma{\bf{k}}}(\nu_{n}),j_{\tilde{b}\sigma{\bf{k}}}(\nu_{n})\right]^{T}.
   	\nonumber\\
   	\label{Equation_12}
   	\end{eqnarray}
    	Here, we use the condition of the half-filling in each layer of the graphene bilayer, in order to determine the exact chemical potential $\mu$ in the interacting system. For the layer 1, this condition holds that 
   \begin{eqnarray}
   \left\langle{n}_{a}\right\rangle+\left\langle{n}_{b}\right\rangle=1.
   \label{Equation_13}
   \end{eqnarray}
   After evaluating the averages in the left-hand side in Eq.(\ref{Equation_13}), we get for the chemical potential the following equation
   \begin{eqnarray}
   \frac{4}{N}\sum_{{\bf{k}}}\sum^{4}_{i=1}\alpha_{i{{\bf{k}}}}n_{\rm F}(\xi_{i{\bf{k}}})=1.
   \label{Equation_14}
   \end{eqnarray}
   The coefficients $\alpha_{i{{\bf{k}}}}$, in Eq.(\ref{Equation_14}) are defined in Eq.(21) in Ref.\onlinecite{cite_46}.
   The function $n_{F}\left(\varepsilon\right)$, in Eq.(\ref{Equation_14}), is the Fermi-Dirac distribution function $n_{F}\left(\varepsilon\right)=1/\left[e^{\beta{\left(\varepsilon-\mu\right)}}+1\right]$.  
   For the parameters $\xi_{i{\bf{k}}}$, in the arguments of the Fermi-Dirac distribution function, in Eq.(\ref{Equation_14}), we have
   \begin{eqnarray}
   \xi_{i{\bf{k}}}=\mu-\kappa_{i{\bf{k}}},
   \label{Equation_15}
   \end{eqnarray}
   where the energy the parameters $\kappa_{i{\bf{k}}}$, define the electronic band structure of the bilayer graphene with the excitonic pairing interaction. We find that
   {
   	
   	\begin{align}
   	\kappa_{1,2{\bf{k}}}=-\frac{1}{2}\left[\Delta+\gamma_{1}\pm\sqrt{\left(W-\Delta-\gamma_{1}\right)^{2}+4|\tilde{\gamma}_{{\bf{k}}}|^{2}}\right]+\bar{\mu},
   	\nonumber\\
   	\kappa_{3,4{\bf{k}}}=-\frac{1}{2}\left[-\Delta-\gamma_{1}\pm\sqrt{\left(W+\Delta+\gamma_{1}\right)^{2}+4|\tilde{\gamma}_{{\bf{k}}}|^{2}}\right]+\bar{\mu}.
   	\nonumber\\
   	\label{Equation_16}
   	\end{align}}
   \newline\\

   Here, we have introduced a new bare chemical potential $\bar{\mu}$
   \begin{eqnarray}
   \bar{\mu}=\frac{\mu^{\rm eff}_{1}+\mu^{\rm eff}_{2}}{2}.
   \label{Equation_17}
   \end{eqnarray}

   	After evaluating the statistical average, given in Eq.(\ref{Equation_3}), we will have the self-consistent equation for the excitonic gap parameter $\Delta$
   	\begin{eqnarray}
   	\Delta=\frac{W(\gamma_{1}+\Delta)}{N}\sum_{{\bf{k}}}\sum^{4}_{i=1}\beta_{i{{\bf{k}}}}n_{F}(\xi_{i{\bf{k}}}).
   	\label{Equation_18}
   	\end{eqnarray}
   Here, the coefficients $\beta_{i{{\bf{k}}}}$ are given in Eq.(26) in Ref.\onlinecite{cite_46}. 
   	
   	As we will see later on, when evaluating numerically $\kappa_{i{\bf{k}}}$, the shift of the Dirac's crossing energy level, due to the non-zero condensate states in the noninteracting BLG (in the case $W=0$ the condensate states in the BLG are due to the interlayer hopping mechanism) is directly related to the effective bare chemical potential $\bar{\mu}$, which enters in Eq.(\ref{Equation_16}). It is not difficult to verify that for the case of the noninteracting BLG, i.e., when $U=0$, $W=0$ and $\Delta=0$, the expressions in Eq.(\ref{Equation_16}) are reducing to the usual tight-binding dispersion relations 
   	
   	\begin{eqnarray}
   	\varepsilon_i=\pm\frac{\gamma_1}{2}\pm \sqrt{({k}\gamma_0)^{2}+(\gamma_1/2)^{2}}-\mu, 
   	\label{Equation_19}
   	\end{eqnarray}
   	with $i=1,..4$ and $k=|\gamma_{{\bf{k}}}|^{2}$) discussed in Ref.\onlinecite{cite_41}, in the context of the real-space Green's function study of the noninteracting bilayer graphene. 
   	
   	The exact numerical solution for the excitonic pairing gap parameter $\Delta$ in the BLG is given in Ref.\onlinecite{cite_46}. As the numerical calculations show, in the mentioned paper, the pairing gap parameter decreases when augmenting the temperature and is not destroyed at the very high temperatures. It is also clear from the results, given in Ref.\onlinecite{cite_46}, that the weak coupling region corresponds to the BCS-like excitonic pairing states between the electrons and holes in different layers of the BLG, while the strong coupling region corresponds to the excitonic condensate states in the system. In the next Section, we will present the modifications of the BLG band structure due to the excitonic effects and we will calculate the hybridization gap $\Delta_H=|\kappa_{2{\bf{k}}}-\kappa_{3{\bf{k}}}|$ at the Dirac's point $K$, between the conduction band electrons and valence band holes in the bilayer graphene.
   	%
   	\subsection{\label{sec:Section_3_2} The excitonic band structure and the hybridization gap}
   	%
   	Here, we will discuss in more details the numerical results for the electronic band structure of the bilayer graphene in the vicinity of the Dirac's point $K$ and by considering the excitonic pairing mechanism in the BLG. The detailed band structure of the BLG with the excitonic pairing mechanism is discussed in Ref.\onlinecite{cite_46}.  
   	   	  
\begin{figure*}  
   	  			\begin{center}
   	  				\includegraphics[width=520px,height=230px]{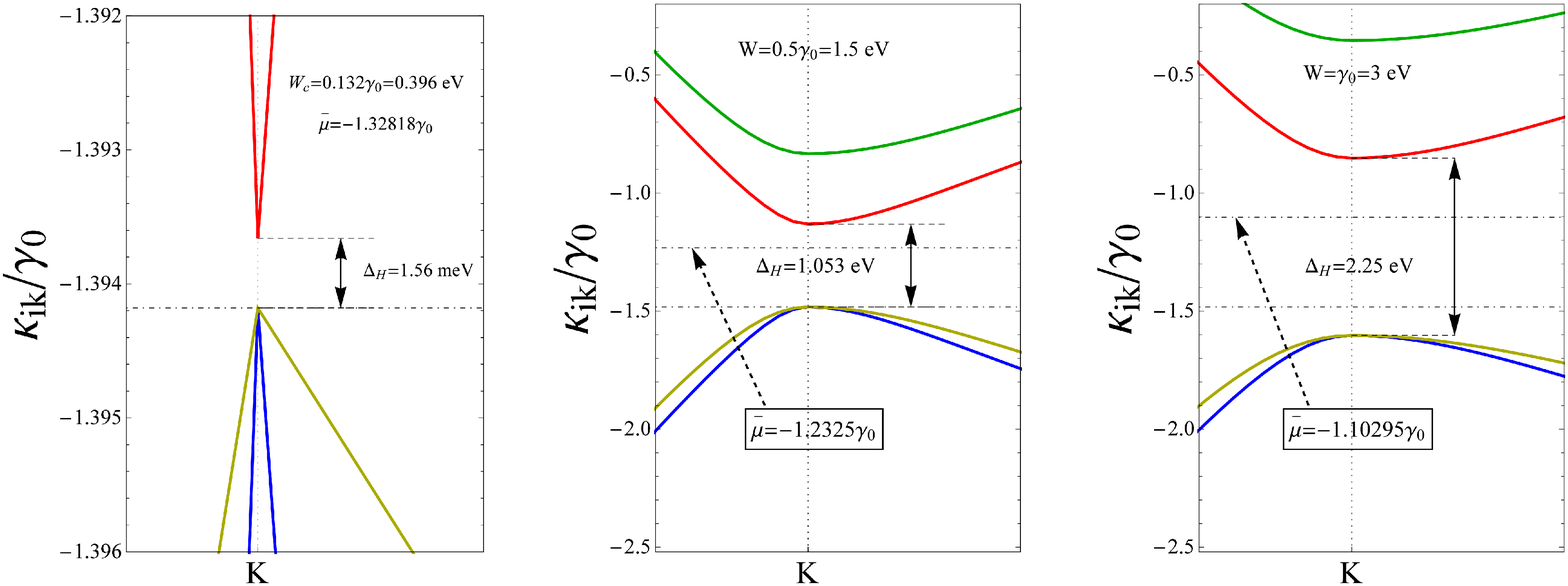}
   	  				\caption{\label{fig:Fig_2}(Color online) The Electronic band structure of the bilayer graphene in the vicinity of the Dirac's point $K$, for various values of the normalized interlayer Coulomb interaction parameter $W/\gamma_0$. The values of the hybridization gap $\Delta_{H}$, and the Fermi energy levels $\bar{\mu}$ are shown in the panels. The interlayer hopping amplitude is fixed at the value $\gamma_1=0.128\gamma_0$, and the zero temperature limit is considered. } 
   	  			\end{center}
   	  		\end{figure*}

\begin{figure*}  
   	  			\begin{center}
   	  				\includegraphics[width=520px,height=230px]{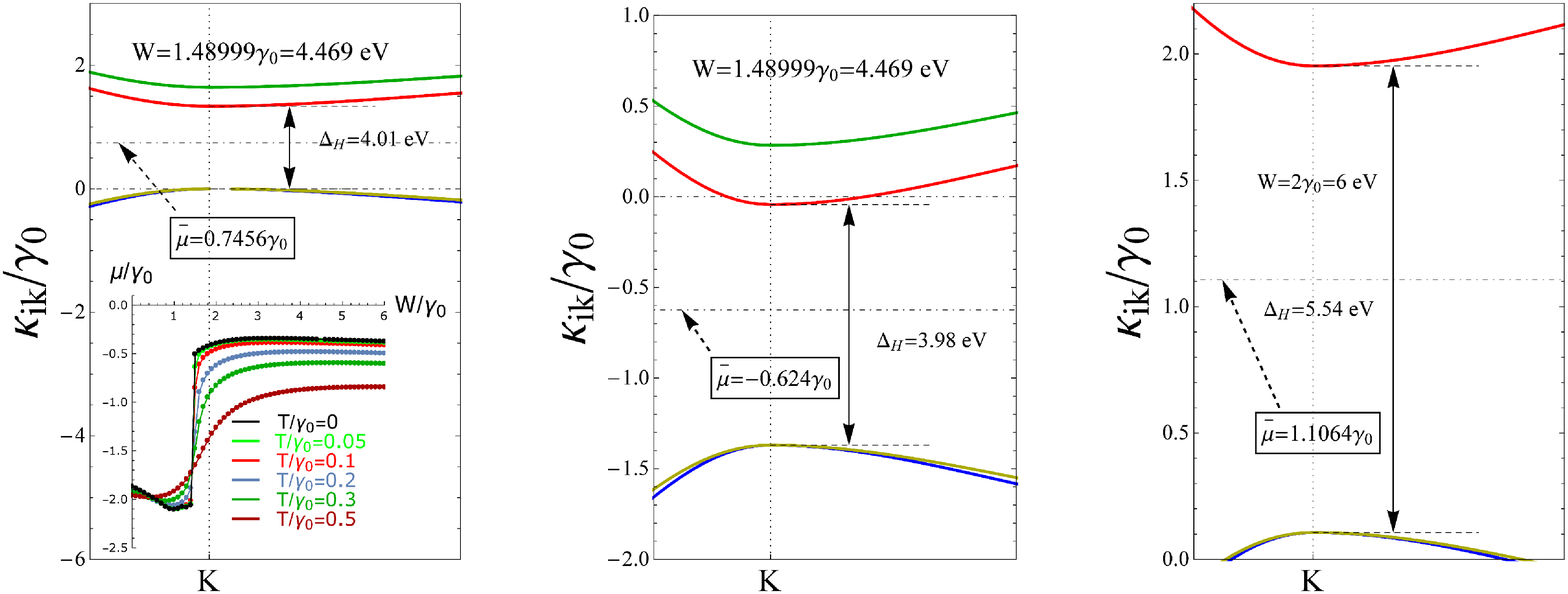}
   	  				\caption{\label{fig:Fig_3}(Color online) The electronic band structure of the bilayer graphene near the Dirac's point $K$, for various values of the normalized interlayer Coulomb interaction parameter $W/\gamma_0$. The values of the hybridization gap $\Delta_{H}$ and the Fermi energy levels $\bar{\mu}$ are shown in the panels. The interlayer hopping amplitude is fixed at the value $\gamma_1=0.128\gamma_0$, and the zero temperature limit is considered. The inset, in the bottom of the first panel, shows the behavior of the chemical potential as a function of the interlayer interaction parameter $W\gamma_0$ and for different values of temperature.} 
   	  			\end{center}
   	  		\end{figure*}

  	In Figs.~\ref{fig:Fig_2} and Fig.~\ref{fig:Fig_3} the electronic band structures (see in Eq.(\ref{Equation_16})) are shown near the Dirac's $K$ point, where the hybridization gap starts to open at the critical value of the interlayer interaction parameter $W_{c}=0.132\gamma_0=0.396$ eV (we have putted here the realistic value for the intralayer hopping parameter $\gamma_0=3$ eV, according to Ref.\onlinecite{cite_42}).  In the inset, in the first panel, in Fig.~\ref{fig:Fig_3}, we have shown the exact solution for the chemical potential $\mu$ in the bilayer graphene as a function of the interlayer Coulomb interaction parameter $W/\gamma_0$. Different values of temperature have been considered there. 
   	%
   	  		%
   	  	%
   	  	
For the noninteracting case ($W=0$), which is given in the first of the left panels, in Fig.~\ref{fig:Fig_3}, we recover the usual BLG band structure modified by the displacement of the Dirac's crossing energy $\epsilon_{{\cal{D}}}$. This is due to the finite chemical potential solution in the interacting system. For the zero interaction limit, the bare chemical potential $\bar{\mu}$ is coinciding exactly with the Dirac's crossing energy level: $\bar{\mu}=\epsilon_{{\cal{D}}}$. The differences appear at the non-zero values of $W$.  Therefore, $\bar{\mu}$ controls the position of the Fermi level in the interacting BLG.

It is calculated using the formula in Eq.(\ref{Equation_17}), in the Section \ref{sec:Section_3_1}: $\bar{\mu}=\mu+\kappa{U}+0.5W$ (with $\kappa=0.25$), where we have used the exact solution for the chemical potential $\mu$, after the self-consistent equations for the excitonic gap parameter $\Delta$ and the chemical potential $\mu$. It has been shown in Ref.\onlinecite{cite_46}, that the Fermi energy in the bilayer graphene at $T=0$ has also a very large jump nearly at the same value of $W$ as the exact chemical potential $\mu$: $W\sim1.49\gamma_0$ (see in the inset, in the first panel, in Fig.~\ref{fig:Fig_3}). In contrast, for higher temperatures $T>0.05\gamma_0$, this behavior is smoothed, and for the very high temperatures, we have practically the continuous variation of $\bar{\mu}$ with respect to $W$. The behavior of the Fermi energy, give in Ref.\onlinecite{cite_46} explains, at least qualitatively, the similar behavior of the bilayer graphene chemical potential, observed experimentally in \cite{cite_38}, by a direct measurement of the chemical potential of the BLG as a function of its carrier density. For this purpose, a double-BLG heterostructure has been built, and the bottom bilayer chemical potential has been mapped along the charge neutrality line of the top bilayer (see in Ref.\cite{cite_38}, for details).
 
In the table \ref{tab:a}, we have presented the exact solutions of the chemical potential $\mu$, bare chemical potential $\bar{\mu}$, and the Dirac's crossing energies $\varepsilon_{{\cal{D}}}$ for all considered values of the parameter $W/\gamma_0$. The levels of the bare chemical potential $\bar{\mu}$, which indeed play the role of the Fermi energy for each given interaction value, are presented in all panels, in Figs.~\ref{fig:Fig_2} and Fig.~\ref{fig:Fig_3}. We see in the first panel in Fig.~\ref{fig:Fig_2} that the hybridization gap $\Delta_{H}$ opens in the electronic band structure of the bilayer graphene at the critical value $W_{c}=0.132\gamma_0=0.396$ eV of the interlayer Coulomb interaction parameter. Nevertheless, the system is formally semi-metallic in this case, because the Fermi level $\bar{\mu}$ lies in the conduction band ($\bar{\mu}=-1.32818\gamma_0$). This is the limit of the small gap semiconductor when the system has the semi-metallic properties yet and at the same time is in the optically active regime. We observe also in Figs.~\ref{fig:Fig_2} and ~\ref{fig:Fig_3} that the hybridization gap increases when augmenting the interaction parameter $W$. The true semiconducting limit happens for the higher values of $W$, which we see in the middle panel in Fig.~\ref{fig:Fig_2}, where the system passes into the semiconducting state at $W=0.5\gamma_0=1.5$ eV, and the Fermi level is situated in the hybridization gap region $\bar{\mu}=-1.2325\gamma_0$. We, intentionally, have not attributed to $\bar{\mu}$ the Fermi energy notation $\epsilon_{{\cal{F}}}$ in the pictures, in order to distinguish it from the noninteracting case. For the larger values of $W$ (see in the panels, in Fig.~\ref{fig:Fig_3}), the Fermi energy lies practically in the middle of the hybridization gap, reporting the well defined intrinsic semiconducting state (or insulator state for the large enough values of $W$) in the bilayer graphene. It is interesting to observe that the band structure of the BLG degenerates at the interlayer interaction value $W_{j}=1.48999\gamma_0$ and $T=0$. Indeed, we see in the first two panels in Fig.~\ref{fig:Fig_3} that there are two band structures for the same value of $W=W_j$. This degeneracy is related to the chemical potential solution in the system. We have two possible chemical potential solutions at $W=1.48999\gamma_0$ (see also in the table \ref{tab:a}). One is situated in the lower bound of the solution for the chemical potential $\mu$ (see in the inset, in the first panel, in Fig.~\ref{fig:Fig_3}) and the another solution is in the upper bound solution of $\mu$. Therefore, the chemical potential has a large jump at $W_{j}=1.48999\gamma_0$, from the lower bound to the upper bound: $\Delta\mu=1.37\gamma_0=4.11$ eV. For the higher values of $W$: $W>W_{j}$, we have only the upper bound solutions for $\mu$. Those mentioned values of $\mu$ at $W=W_{j}$ are given in the table \ref{tab:a}.
   	
\begin{figure}  
   		\begin{center}
   			\includegraphics[width=200px,height=200px]{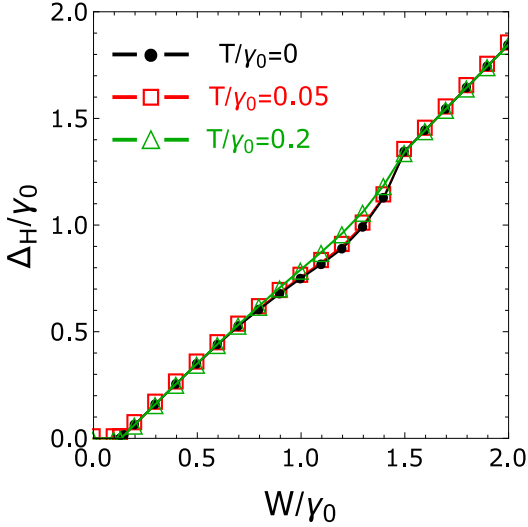}
   			\caption{\label{fig:Fig_4}(Color online) The hybridization gap $\Delta_{H}/\gamma_0$ as a function of the Coulomb interaction parameter $W/\gamma_0$ and for various values of temperature. The interlayer hopping amplitude is fixed at the value $\gamma_1=0.128\gamma_0$.}
   		\end{center}
   	\end{figure}

It is worth to mention that the gap opening in the bilayer graphene's band structure at the Dirac's point $K$, discussed here, has not its analogs in the literature yet. Particularly, it is not equivalent to the studies when the external electric field creates a tunable band gap in the band structure of the bilayer graphene (see, for example in Refs.\onlinecite{cite_2,cite_3}).   
  	\begin{table*}[t]
   		\centering
   		\begin{tabular}{{c|c|c|c|c|c|c|c|c}}
   			$W$ & 0 & 0.12$\gamma_0$ & 0.132$\gamma_0$ & 0.5$\gamma_0$ & $\gamma_0$ & 1.48999$\gamma_0$ (lower bound)  & 1.48999$\gamma_0$ (upper bound) & 2$\gamma_0$\\
   			\hline
   			\\
   			$\mu$ & -1.863$\gamma_0$ & -1.89$\gamma_0$  & -1.894$\gamma_0$ & -1.982$\gamma_0$  & -2.103$\gamma_0$ & -1.86956$\gamma_0$  & -0.49935$\gamma_0$ & -0.393$\gamma_0$\\
   			$\bar{\mu}$ & -1.363$\gamma_0$ & -1.3306$\gamma_0$ & -1.328$\gamma_0$ & -1.2325$\gamma_0$  & -1.103$\gamma_0$ & -0.624$\gamma_0$ & 0.7456$\gamma_0$ & 1.1064$\gamma_0$\\
   			${\varepsilon}_{{\cal{D}}}$ & -1.363$\gamma_0$ & -1.3907$\gamma_0$ & -1.39418$\gamma_0$ & -1.4825$\gamma_0$ & -1.603$\gamma_0$ & -1.369$\gamma_0$ & 0.00065$\gamma_0$ & 0.106$\gamma_0$ \\
   			$\Delta_{H}$ & 0 & 0 & 0.00052$\gamma_0$ & 0.351$\gamma_0$ & 0.75$\gamma_0$ & 1.326$\gamma_0$ & 1.336$\gamma_0$ & 1.846$\gamma_0$ \\
   		\end{tabular}
   		\caption{The exact solutions of the chemical potential $\mu$, Fermi energy $\bar{\mu}$, Dirac's crossing energy $\varepsilon_{\cal{D}}$ and the hybridization gap $\Delta_H$.}
   		\label{tab:a}
   	\end{table*}   		
In Fig.~\ref{fig:Fig_4}, we have presented the dependence of the hybridization gap on the interlayer interaction parameter $W$. Here, the hybridization gap is due to the local interlayer Coulomb interaction effects, which is not equivalent to the electric field effects, discussed in Refs.\cite{cite_2,cite_3}. Three different limits of temperature have been considered in Fig.~\ref{fig:Fig_4}. Nearly linear dependence of $\Delta_{H}$ on $W/\gamma_0$, for the small and intermediate values of $W$, remains practically unchanged when varying the temperature. This linear behavior corresponds to the decreasing values of the chemical potential solution in the interval $W\in(0, \gamma_0)$ (see the lower bound solution for the chemical potential, in the inset, in the first panel, in Fig.~\ref{fig:Fig_3}. We observe, in Fig.~\ref{fig:Fig_4}, that the gap formation in the bilayer graphene is truly a threshold process, and for $W<W_c=0.132\gamma_0$ we have $\Delta_H\equiv0$.  At $W=0.132\gamma_0=0.396$ eV the Fermi energy level $\bar{\mu}$ is situated in the conduction band, nevertheless the hybridization gap is not zero: $\Delta_H=1.56$ meV. The threshold value of the interlayer interaction parameter at which the hybridization gap opens at $T=0$ is $W=0.132\gamma_0=0.396$ eV, corresponding to $\Delta_{H}=0.000517\gamma_0=1.551$ meV. For $T=0.05\gamma_0$, we have $\Delta_{H}=0.000441\gamma_0=1.32$ meV, corresponding to $W=0.132\gamma_0=0.3945$ eV, and for  
$T=0.2\gamma_0$: $\Delta_{H}=0.00037\gamma_0=1.11$ meV, corresponding again to $W=0.132\gamma_0=0.396$ eV. We see that $\Delta_{H}$ is slightly decreasing when augmenting the temperature, while the critical value of the interlayer interaction parameter $W_{c}$ remains the same: $W_{c}=0.132\gamma_0=0.396$ eV. The deviations from the linearity of $\Delta_{H}$ in the interval $W\in\left[1.1\gamma_0, 1.49\gamma_0\right)$ is related again to the behavior of the chemical potential solution in this interval (see in the inset, in the first panel, in Fig.~\ref{fig:Fig_3}, black line). Indeed, the non-linearity of $\Delta_{H}$ could not be explained with the behavior $\bar{\mu}$, nevertheless $\bar{\mu}$ plays the role of the Fermi level in the interacting bilayer graphene (see in the Section 5, in Ref.\onlinecite{cite_46}, the discussion on the parameter $\bar{\mu}$). The bare chemical potential $\bar{\mu}$, obtained in the paper is the effective chemical potential, which is directly related to the measurements of the Fermi level in the bilayer graphene \cite{cite_38}, while the original energy parameter $\mu$ is the prerequisite in order to explain the behavior of the hybridization gap. The second linear part of the hybridization function curves (see the interval of the interaction parameter $W\in\left[1.49\gamma_0, 2\gamma_0\right]$) corresponds to the upper bound solutions of the chemical potential in the bilayer graphene system. 
%
   	%
   	
   	\section{\label{sec:Section_4} The single-particle spectral properties}
   	%
   	\subsection{\label{sec:Section_4_1}  The sublattice spectral functions}
   	%
   	We will introduce here the normal Matsubara Green's functions, according to the standard notations \cite{cite_36, cite_37}, namely, for the sublattice-$\eta$, in the layer $\ell$, we define the normal local Green's functions, as follows
   	\begin{eqnarray}
   	G_{\eta\sigma}\left({\bf{r}}\tau,{\bf{r}}\tau\right)=\left\langle \eta_{\sigma}({\bf{r}}\tau){\eta}^{\dag}_{\sigma}({\bf{r}}\tau)\right\rangle.
   	\label{Equation_20}
   	\end{eqnarray}
   	In the Fourier space representation, and for the sublattice-$A$ in the bottom layer 1, the Green's function, in Eq.(\ref{Equation_20}), takes the following form
   	\begin{eqnarray}
   	G_{a\sigma}\left({\bf{r}}\tau,{\bf{r}}\tau\right)=\frac{1}{\beta{N}}\sum_{{\bf{k}}\nu_{n}}G_{a\sigma}\left({\bf{k}},\nu_{n}\right),
   	\label{Equation_21}
   	\end{eqnarray}
   	where the Fourier transform $G_{a\sigma}\left({\bf{k}},\nu_{n}\right)$ is given as
   	\begin{eqnarray}
   	G_{a}\left({\bf{k}},\nu_{n}\right)=\frac{1}{\beta{N}}\left\langle a_{{\bf{k}}}(\nu_{n}){a}^{\dag}_{{\bf{k}}}(\nu_{n})\right\rangle.
   	\label{Equation_22}
   	\end{eqnarray}
   	The function $G_{a}\left({\bf{k}},\nu_{n}\right)$, in Eq.(\ref{Equation_22}), is the Matsubara Green's function for the $\eta=a$-type fermions in the layer 1.
   	Here, we have restricted only to the case $\sigma=\uparrow$ and we have omitted the spin indexes in the Green's functions notations. This is due to the spin-symmetry of the action, given in Eq.(\ref{Equation_4}). The statistical average in the expression of the Green's function $G_{a}\left({\bf{k}},\nu_{n}\right)$ could be evaluated with the help of the partition function, given in Eq.(\ref{Equation_11}), in the Section \ref{sec:Section_3_1}.
   	
   	The normal spectral functions for different sublattices (corresponding to the sites $A$ and $B$), in the layer 1, are defined on the real frequency axis $\nu$ with the help of the retarded Green's functions \cite{cite_37}. For the sublattice-$A$, we have
   	\begin{eqnarray}
   	{\cal{S}}_{a}({\bf{k}},\nu)=-\frac{1}{\pi}\Im{G^{\rm R}_{a}({\bf{k}},\nu)}.
   	\label{Equation_23}
   	\end{eqnarray} 
   	The retarded Green's function, in turn, could be obtained after the analytical continuation of the corresponding Matsubara Green's function $G_{a}({\bf{k}},\nu_{n})$, given in Eq.(\ref{Equation_22}) above:
   	\begin{eqnarray}
   	G^{\rm R}_{a}({\bf{k}},\nu)=\left.G_{a}({\bf{k}},\nu_{n})\right\vert_{i\nu_{n}\rightarrow \nu+i0^{+}}.
   	\label{Equation_24}
   	\end{eqnarray} 
   	The explicit form of the Green's function $G_{a}({{\bf{k}}},\nu_{n})$ could be obtained after the functional derivation with respect to the external source field variables and by using the partition function, in Eq.(\ref{Equation_10}). For the sublattice-$A$, in the layer 1 of the BLG, we get
   	\begin{eqnarray}
   	G_{a}({\bf{k}},\nu_{n})=\sum^{4}_{i=1}\frac{\alpha_{i{\bf{k}}}}{i\nu_{n}+\kappa_{i{\bf{k}}}}.
   	\label{Equation_25}
   	\end{eqnarray}
   	Then, for the single-particle spectral function ${\cal{S}}_{a}({\bf{k}},\nu)$, we get the following expression 
   	\begin{eqnarray}
   	{\cal{S}}_{a}({\bf{k}},\nu)=\frac{1}{\pi}\sum^{4}_{i=1}\frac{\lambda\alpha_{i{\bf{k}}}}{\lambda^{2}+\left(\nu+\kappa_{i{\bf{k}}}\right)^{2}}.
   	\label{Equation_26}
   	\end{eqnarray}
   	We have introduced the Lorentzian function representation for the Dirac's delta-function, and a broadening parameter $\lambda$ has been introduced in Eq.(\ref{Equation_26}). Furthermore, we calculate the normal spectral function for the sublattice-$B$, in the layer 1. Similarly, we obtain  
   	\begin{eqnarray}
   	{\cal{S}}_{b}({\bf{k}},\nu)=\frac{1}{\pi}\sum^{4}_{i=1}\frac{\lambda\gamma_{i{\bf{k}}}}{\lambda^{2}+\left(\nu+\kappa_{i{\bf{k}}}\right)^{2}}
   	\label{Equation_27}
   	\end{eqnarray}
   	with
   	\begin{align}
   	\footnotesize
   	\arraycolsep=0pt
   	\medmuskip = 0mu
   	\gamma_{i{{\bf{k}}}}
   	=(-1)^{i+1}
   	\left\{
   	\begin{array}{cc}
   	\displaystyle  & \prod^{}_{j=3,4}\frac{{\cal{P}'}^{(3)}(\kappa_{i{\bf{k}}})}{\left(\kappa_{1{\bf{k}}}-\kappa_{2{\bf{k}}}\right)}\frac{1}{\left(\kappa_{i{\bf{k}}}-\kappa_{j{\bf{k}}}\right)},  \ \ \  $if$ \ \ \ i=1,2,
   	\newline\\
   	\newline\\
   	\displaystyle  & \prod^{}_{j=1,2}\frac{{\cal{P}'}^{(3)}(\kappa_{i{\bf{k}}})}{\left(\kappa_{3{\bf{k}}}-\kappa_{4{\bf{k}}}\right)}\frac{1}{\left(\kappa_{i{\bf{k}}}-\kappa_{j{\bf{k}}}\right)},  \ \ \  $if$ \ \ \ i=3,4,
   	\end{array}\right.
   	\nonumber\\
   	\label{Equation_28}
   	\end{align}
   	where ${\cal{P}'}^{(3)}(\kappa_{i{\bf{k}}})$ is the polynomial of third order in $\kappa_{i{\bf{k}}}$, namely we have
   	\begin{eqnarray}	{\cal{P}'}^{(3)}(\kappa_{i{\bf{k}}})=\kappa^{3}_{i{\bf{k}}}+\omega'_{1{\bf{k}}}\kappa^{2}_{i{\bf{k}}}+\omega'_{2{\bf{k}}}\kappa_{i{\bf{k}}}+\omega'_{3\bf{k}}
   	\label{Equation_29}
   	\end{eqnarray}
   	with the coefficients $\omega'_{i{\bf{k}}}$, $i=1,...3$, given as 
   	\begin{align}
   	&&\omega'_{1{\bf{k}}}=-2\mu^{\rm eff}_{1}-\mu^{\rm eff}_{2},
   	\nonumber\\
   	&&\omega'_{2{\bf{k}}}=\mu^{\rm eff}_{1}\left(\mu^{\rm eff}_{1}+2\mu^{\rm eff}_{2}\right)-|\tilde{\gamma}_{{\bf{k}}}|^{2},
   \nonumber\\
   	&&\omega'_{3{\bf{k}}}=-\mu^{\rm eff}_{2}\left(\mu^{\rm eff}_{1}\right)^{2}+\mu^{\rm eff}_{1}|\tilde{\gamma}_{{\bf{k}}}|^{2}.
   	\label{Equation_30}
   	\end{align}
   	%
   	\subsection{\label{sec:Section_4_2} The anomalous spectral function and the excitonic coherence length}
   	%
   	Another interesting function, to be considered here, is the anomalous spectral function $S_{b\tilde{a}}$, defined between the layers of the BLG, which (theoretically) gives a direct information about the excitonic pair formation and condensation in the double layer graphene system. This function provides the simultaneous probability to find an electron in the upper layer 2 and a hole in the lower layer 1, for a given discrete quantum state $({\bf{k}},\nu)$, thus, reporting directly the excitonic pairs between the layers of the bilayer graphene. The frequency integrated expression of this function is especially relevant, due to its direct relation to the excitonic coherence length $\xi_c$ between the layers. The spectral function $S_{b\tilde{a}}$ could be obtained with the help of the imaginary part of the anomalous retarded Green's function $G^{\rm R}_{b\tilde{a}}({\bf{k}},\nu)$. We have
   	\begin{eqnarray}
   	G^{\rm R}_{b\tilde{a}}({\bf{k}},\nu)=\left.G_{b\tilde{a}}({\bf{k}},\nu_{n})\right\vert_{i\nu_{n}\rightarrow \nu+i0^{+}},
   	\label{Equation_31}
   	\end{eqnarray}
   	where $G_{b\tilde{a}}({\bf{k}},\nu_{n})$ is the Fourier transform of the anomalous Matsubara Green's function. It is defined as follows:
   	\begin{eqnarray}
   	G_{b\tilde{a}}({\bf{k}},\nu_{n})=\frac{1}{\beta{N}}\left\langle \tilde{a}({\bf{k}},\nu_{n}){b}^{\dag}({\bf{k}},\nu_{n})\right\rangle.
   	\label{Equation_32}
   	\end{eqnarray}
   	Then, we have the anomalous spectral function
   	\begin{eqnarray}
   	{\cal{S}}_{b\tilde{a}}({\bf{k}},\nu)=-\frac{1}{\pi}\Im{G^{\rm R}_{b\tilde{a}}({\bf{k}},\nu)}.
   	\label{Equation_33}
   	\end{eqnarray}
   	Again, after using the functional derivation techniques, we get the following analytical form
   	\begin{eqnarray}
   	{\cal{S}}_{b\tilde{a}}({\bf{k}},\nu)=\frac{\gamma_{1}+\Delta}{\pi}\sum^{4}_{i=1}\frac{\lambda\alpha_{i{\bf{k}}}}{\lambda^{2}+\left(\nu+\kappa_{i{\bf{k}}}\right)^{2}}.
   	\label{Equation_34}
   	\end{eqnarray}
   	It is not difficult to show that $G_{22}({\bf{k}},\nu_{n})=G_{33}({\bf{k}},\nu_{n})$ and $G_{11}({\bf{k}},\nu_{n})=G_{44}({\bf{k}},\nu_{n})$, which follows from the symmetry of the inverse Green's function matrix, given in Eq.(\ref{Equation_6}). Therefore, we have for the spectral functions
   	\begin{eqnarray}
   	{\cal{S}}_{\tilde{a}}({\bf{k}},\nu)&=&{\cal{S}}_{b}({\bf{k}},\nu),
   	\nonumber\\
   	{\cal{S}}_{\tilde{b}}({\bf{k}},\nu)&=&{\cal{S}}_{a}({\bf{k}},\nu).
   	\label{Equation_35}
   	\end{eqnarray}
   	The relations in Eq.(\ref{Equation_35}) above explicitly imply that the spectral functions, for the layer 2 in the BLG, coincide with that of the layer 1, just after interchanging the sublattice sites. After the summation over the Matsubara frequencies $\nu_{n}$ in the expressions of the normal and anomalous Green's functions, we will get the momentum dependent correlation functions of the system. Namely, the frequency dependent Matsubara Green's functions in the BLG have the following structures
   	\begin{eqnarray}
   	G_a({\bf{k}},\nu_{n})=\sum^{4}_{i=1}\frac{\alpha_{i{\bf{k}}}}{i\nu_{n}+\kappa_{i{\bf{k}}}},
\nonumber\\
	G_{b\tilde{a}}({\bf{k}},\nu_{n})=\Delta'\sum^{4}_{i=1}\frac{\beta_{i{\bf{k}}}}{i\nu_{n}+\kappa_{i{\bf{k}}}},
\nonumber\\
   	G_b({\bf{k}},\nu_{n})=\sum^{4}_{i=1}\frac{\gamma_{i{\bf{k}}}}{i\nu_{n}+\kappa_{i{\bf{k}}}},
   	\label{Equation_36}
   	\end{eqnarray}
   	where $\Delta'$ in the last equation is $\Delta'=\Delta+\gamma_1$.
   	After summing over the fermionic Matsubara frequencies $\nu_{n}$ $g_{\eta}({\bf{k}})=\frac{1}{\beta}\sum_{\nu_{n}}G_{\eta}({\bf{k}},\nu_{n})$, we get the momentum distribution functions in the BLG  
   	\begin{eqnarray}
   	g_a({\bf{k}})=\sum^{4}_{i=1}{\alpha_{i{\bf{k}}}}n_{F}(-\kappa_{i{\bf{k}}}),
\nonumber\\
   	g_{b\tilde{a}}({\bf{k}})=\Delta'\sum^{4}_{i=1}{\beta_{i{\bf{k}}}}n_{F}(-\kappa_{i{\bf{k}}}),
   	\nonumber\\
   	g_b({\bf{k}})=\sum^{4}_{i=1}{\gamma_{i{\bf{k}}}}n_{F}(-\kappa_{i{\bf{k}}}).
   	\label{Equation_37}
   	\end{eqnarray}
   	
\begin{figure}  
   		\begin{center}
   			\includegraphics[width=200px,height=200px]{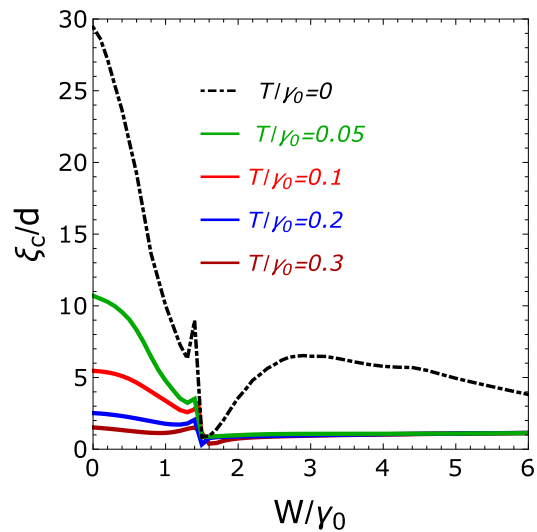}
   			\caption{\label{fig:Fig_5}(Color online) The interlayer exciton coherence length as a function of the interlayer Coulomb interaction parameter $W/\gamma_0$. Different values of temperature are considered in the picture ($T=0$, $T=0.05\gamma_0$, $T=0.1\gamma_0$, $T=0.2\gamma_0$ and $T=0.3\gamma_0$, from top to bottom). The interlayer hopping amplitude is set at the value $\gamma_1=0.128\gamma_0$.}
   		\end{center}
   	\end{figure}

   	In order to see the character of the exciton condensation in the momentum space, we have considered the anomalous momentum distribution function $g_{b\tilde{a}}({\bf{k}})$ (see the second equation in Eq.(\ref{Equation_37})) (see also in the Sections \ref{sec:Section_4_3} and \ref{sec:Section_4_4} below). This function is the exciton condensation amplitude. Note that we  use here the term ``anomalous'' to indicate that the number of electrons on each of the ``$a$'' and ``$b$'' flavours is not conserved due to the excitonic condensation, although the total number of electrons in a given layer $\ell$ is conserved. By using the expression of the function $g_{b\tilde{a}}({\bf{k}})$, we will evaluate the pair coherence length $\xi_c$, which corresponds to the interlayer spatial size of the electron-hole pair and might be defined by the relation
   	\begin{eqnarray}
   	\xi_c=\sqrt{\sum_{\bf{k}}|\nabla_{\bf{k}}{g_{b\tilde{a}}({\bf{k}})}|^{2}/\sum_{\bf{k}}|{g_{b\tilde{a}}({\bf{k}})}|^{2}}.
   	\label{Equation_38}
   	\end{eqnarray}
   	In Fig.~\ref{fig:Fig_5}, we have presented the coherence length in the bilayer graphene as a function of the normalized interlayer Coulomb interaction parameter $W/\gamma_0$. The special numerical package FADBAD++ \cite{cite_43} is used for the numerical differentiation, which implements the forward automatic differentiation of the condensate amplitude function $g_{b\tilde{a}}(\bf{k})$. Different values of temperature are considered in the picture. The interlayer hopping amplitude is fixed at the value $\gamma_1=0.128\gamma_0=0.384$ eV. We see in Fig.~\ref{fig:Fig_5}, that at the zero value of $W$, the interlayer coherence length is very large and it decreases when augmenting the interaction parameter. 
   	
   	The sadden jump of the coherence length at $T=0$ (and also for higher temperatures) in the interval $W\in(1.4\gamma_0, 1.5\gamma_0)$ is related to the behaviour of the chemical potential solution, presented in the inset, in the first panel, in Fig.~\ref{fig:Fig_3} (see the drastic jump of $\mu$ in the case $T=0$). In turn, the Fermi energy $\bar{\mu}$, also has a large jump in that interval of $W$ and at $T=0$ (see in Ref.\onlinecite{cite_46}). For higher values of $T$, this behavior is smoothed. At the jump point of the chemical potential, which corresponds to the value $W_j=1.48999\gamma_0$ (see the band structure in the first two panels in Fig.~\ref{fig:Fig_3}), the coherence length $\xi_c$ becomes comparable to the inter-atomic distances $d$ in the single-layer graphene in the BLG. It is remarkable to indicate that for the case $T=0$ the coherence length is very large for the values $W>W_j$ (for example, $\xi_c\sim 6.5d$ at the value $W=2.9\gamma_0=8.7$ eV). The very large scale of $\xi_c$ at the weak interaction limit is the manifestation of the excitonic BCS-like pairing, and the excitonic gap parameter is very small in this case (see in  Fig.4, in Ref.\onlinecite{cite_46}). Indeed, in the BCS-like weak coupling region (this corresponds to the interval $W\in(0.0,1.3\gamma_0)$), the excitonic coherence length $\xi_c$ and the excitonic gap parameter $\Delta$ satisfy well the BCS-like relation $\xi_c\sim \hbar{v_{{\cal{F}}}}/\Delta$. On the other hand, in the large momentum approximation, where the spectrum of the BLG is nearly linear \cite{cite_44, cite_46}, we have for the Fermi velocity: $v_{{\cal{F}}}=|\bar{\mu}|/(\hbar{k_{\cal{F}}})$, where $k_{\cal{F}}$ is the Fermi wave vector at the Dirac's points $K$ and $K'$. After analysing the solutions for the excitonic gap parameter $\Delta$ and the bare chemical potential $\bar{\mu}$, for the case $T=0$, we find that the relation between the coherence length and excitonic gap parameter in our case reads as $\xi_{c}/d=\alpha|\bar{\mu}|/({k_{\cal{F}}}\Delta)$, where $\alpha$ is the coefficient showing the deviations from the exact BCS relation above. 
   	
   	As it was mentioned above, the bare average chemical potential $\bar{\mu}$ plays the role of the effective Fermi level in the interacting bilayer graphene: $\bar{\mu}=\epsilon_{{\cal{F}}}$. Thus, the BCS-like relation for the coherence length is well satisfied in the case of the weak and intermediate couplings. Contrary, at the high interlayer coupling limit, i.e., when $W>1.5\gamma_0$, the excitonic coherence length becomes proportional to the excitonic gap parameter $\xi\sim \Delta$ and this corresponds to the excitonic condensate regime in the semiconducting phase. We see in Fig.~\ref{fig:Fig_5} that in the BCS regime the excitonic coherence length could be very large, even at the high temperatures (see, for example, $\xi_c$ at $T=0.1\gamma_0$ in Fig.~\ref{fig:Fig_5}, the red curve). Contrary, in the case of strong interaction regime, the excitonic coherence length in the condensate regime is large (of the order of few inter-atomic distances) only in the case $T=0$. For the non-zero temperatures, the coherence length becomes of the order of $d$ in the semiconductor phase. This corresponds to the strong localization of the excitonic states at the surfaces of the BLG layers.  
   	%
   	%
   	\subsubsection{\label{sec:Section_4_3} Zero interlayer interaction}
   	%
   	\begin{figure}  
   		\begin{center}
   			\includegraphics[width=180px,height=160px]{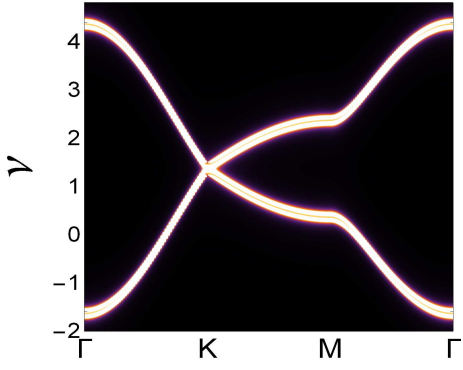}
   			\caption{\label{fig:Fig_6}(Color online) Two-dimensional intensity plot of the $A$-sublattice normal spectral function ${\cal{S}}_{a}({\bf{k}},\nu)$. The interlayer Coulomb interaction parameter is set at zero and the zero temperature limit is considered in the picture. The interlayer hopping amplitude is set at the value $\gamma_1/\gamma_0=0.128$.}
   		\end{center}
   	\end{figure} 
   	Here, we consider the case of the zero interlayer Coulomb interaction limit, i.e. when $W/\gamma_0=0$. The excitonic gap parameter vanishes, meanwhile, the bilayer graphene is strongly correlated in that case (see the discussion below). In Fig.~\ref{fig:Fig_6}, we have plotted the wave vector-resolved intensity plot of the normal spectral function ${\cal{S}}_{a}({\bf{k}},\nu)$ for the zero interlayer Coulomb interaction regime. The chemical potential is finite and negative in that case: $\mu=-1.863\gamma_0$. The temperature, in Fig.~\ref{fig:Fig_6}, is set at zero, and the interlayer hopping is $\gamma_1=0.128\gamma_0$. For the Lorentzian broadening small parameter $\lambda$, we have chosen a very small value $\lambda=0.005$, which is well approximating the Dirac's peaks in the single-quasiparticle spectrum and at the given symmetry points in the Brillouin Zone (BZ).
   	
   	%
   	\begin{figure}  
   		\begin{center}
   			\includegraphics[width=180px,height=160px]{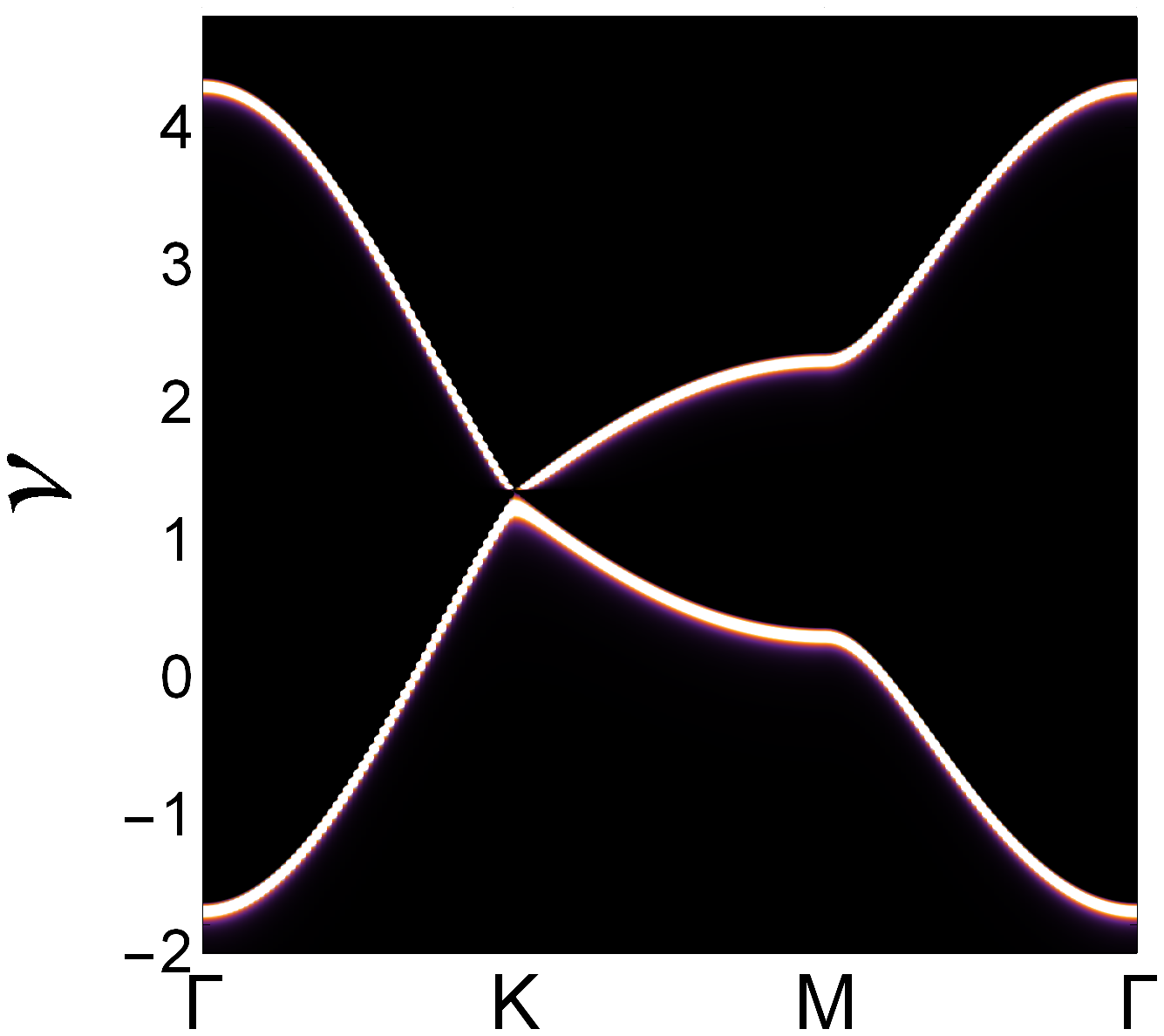}
   			\caption{\label{fig:Fig_7}(Color online) Two-dimensional intensity plot of the anomalous spectral function ${\cal{S}}_{b\tilde{a}}({\bf{k}},\nu)$. The interlayer Coulomb interaction parameter is set at zero, and the zero temperature limit is considered. The interlayer hopping amplitude is set at the value $\gamma_1=0.128\gamma_0$.}
   		\end{center}
   	\end{figure} 
      As we see, there is no gap in the spectral function behaviour at the Dirac neutrality points $K$. Let's mention here the role of the bare chemical potential $\bar{\mu}$ in the normal spectral function structure. We see in Fig.~\ref{fig:Fig_6} that the spectral lines are touching each other at the Dirac's crossing points $K$ and $K'$, as it should be for the case of the noninteracting BLG system \cite{cite_2, cite_3}. The Dirac's crossing energy $\varepsilon_{{\cal{D}}}$ is coinciding exactly with the absolute value of the effective bare chemical potential $\bar{\mu}$: $|\bar{\mu}|=1.363\gamma_0=4.089 $ eV. This value is much higher than the known results for the undoped neutral BLG \cite{cite_45} (where $\varepsilon_{F}=\varepsilon_{D}\sim -\gamma_1\sim -0.4$ eV. This fact is related to the presence of strong correlation effects in the BLG, despite the absence of the interlayer Coulomb interaction. Indeed, the Fermi velocity, calculated at the Dirac's crossing points, has been shown \cite{cite_46} to be much smaller than the Fermi velocity in the non-interacting in-plane graphene sheets \cite{cite_40, cite_47}, and also much smaller than the Fermi velocity of the gated BLG system \cite{cite_38,cite_48} (with $v_F = 1.1\times 10^8 cm\cdot s^{-1}$). This last effect is related to the electron-hole condensate states with a zero center of mass momentum ${\bf{P}}_{c}=0$, at the Dirac's crossing points $K$ and $K'$, and between the opposite layers in the BLG \cite{cite_46}. 
      It is remarkable to note (this will be clear hereafter when discussing the anomalous spectral functions) that the condensate states at zero interlayer interaction turn into the BCS-like weak coupling pairing states when $W/\gamma_0\neq 0$, with a finite excitonic pairing gap $\Delta/\gamma_0\neq 0$ opening in the BLG. 
   	%
   	\begin{figure}  
   		\begin{center}
   			\includegraphics[width=180px,height=160px]{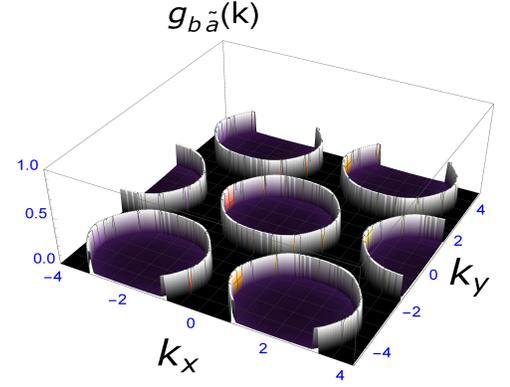}
   			\caption{\label{fig:Fig_8}(Color online) The anomalous spectral function (the condensation amplitude $g_{b\tilde{a}}({\bf{k}})$). The interlayer Coulomb interaction parameter is set at zero, and the zero temperature limit is considered in the picture. The interlayer hopping amplitude is set at the value $\gamma_1=0.128\gamma_0$.}
   		\end{center}
   	\end{figure} 
   	%
   	%
   	\begin{figure}  
   		\begin{center}
   			\includegraphics[width=180px,height=160px]{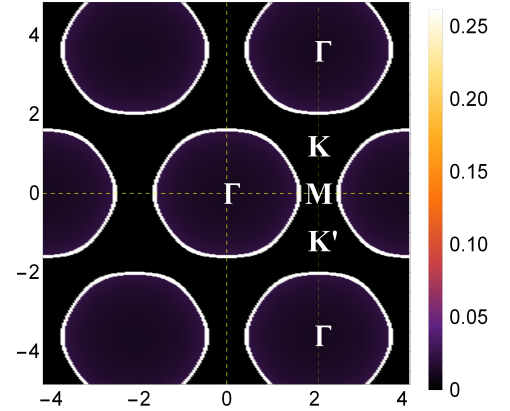}
   			\caption{\label{fig:Fig_9}(Color online) Two-dimensional $(k_{x},k_{y})$ map of the momentum distribution function $g_{b\tilde{a}}({\bf{k}})$. The interlayer Coulomb interaction parameter is set at zero, and the zero temperature limit is considered in the picture. The interlayer hopping amplitude is set at the value $\gamma_1=0.128\gamma_0$.}
   		\end{center}
   	\end{figure} 
   	%
   	\begin{figure}  
   		\begin{center}
   			\includegraphics[width=180px,height=160px]{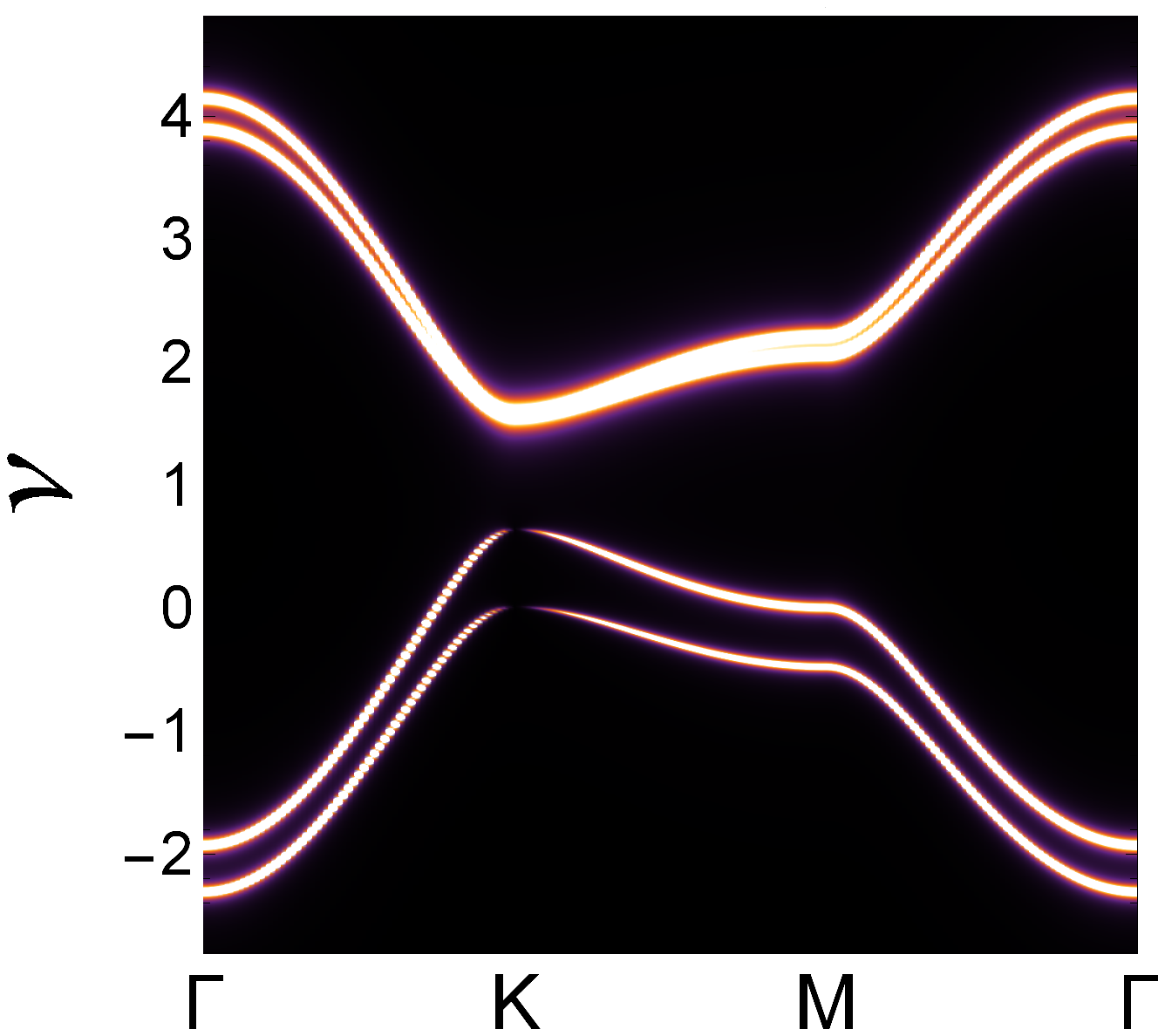}
   			\caption{\label{fig:Fig_10}(Color online) Two-dimensional intensity plot of the $A$-sublattice normal spectral function ${\cal{S}}_{{a}}({\bf{k}},\nu)$. The interlayer Coulomb interaction parameter is set at the value $W=1.25\gamma_0$, and the zero temperature limit is considered in the picture. The interlayer hopping amplitude is set at the value $\gamma_1=0.128\gamma_0$.}
   		\end{center}
   	\end{figure} 
   	%
   	\begin{figure}  
   		\begin{center}
   			\includegraphics[width=180px,height=160px]{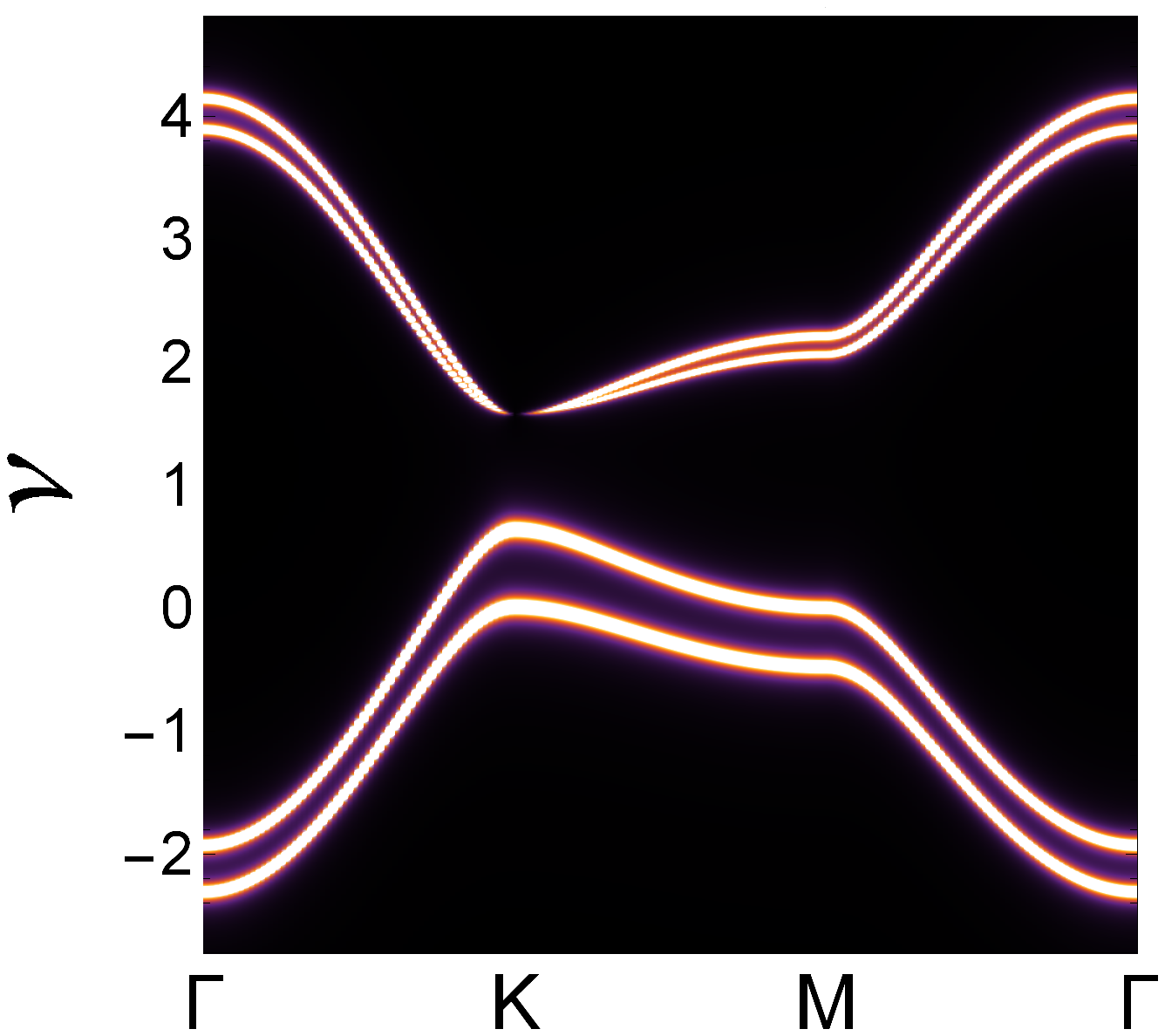}
   			\caption{\label{fig:Fig_11}(Color online) Two-dimensional intensity plot of the $B$-sublattice normal spectral function ${\cal{S}}_{{b}}({\bf{k}},\nu)$. The interlayer Coulomb interaction parameter is set at the value $W=1.25\gamma_0$, and the zero temperature limit is considered in the picture. The interlayer hopping amplitude is set at the value $\gamma_1=0.128\gamma_0$.}
   		\end{center}
   	\end{figure} 
   	%
   	\begin{figure}  
   		\begin{center}
   			\includegraphics[width=180px,height=160px]{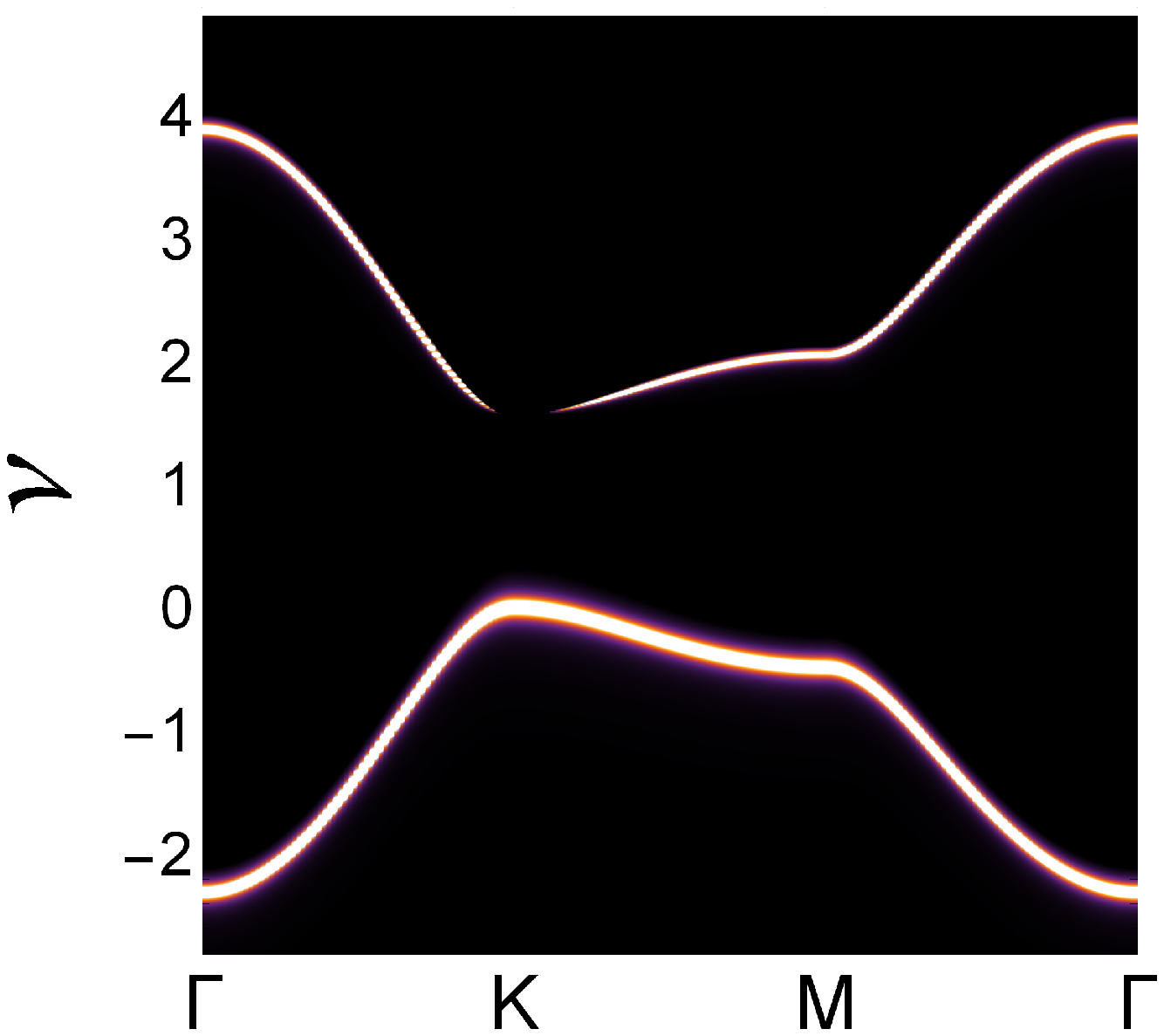}
   			\caption{\label{fig:Fig_12}(Color online) Two-dimensional intensity plot of the anomalous spectral function ${\cal{S}}_{b\tilde{a}}({\bf{k}},\nu)$. The interlayer Coulomb interaction parameter is set at the value $W=1.25\gamma_0$, and the zero temperature limit is considered in the picture. The interlayer hopping amplitude is set at the value $\gamma_1=0.128\gamma_0$.}
   		\end{center}
   	\end{figure} 

In Fig.~\ref{fig:Fig_7}, the anomalous spectral function ${\cal{S}}_{b\tilde{a}}({\bf{k}},\nu)$ is presented for the same values of parameters, as in the case, given in Fig.~\ref{fig:Fig_6}. The spectrum is again gapless at the Dirac's points, and the principal change, compared with the result, presented in Fig.~\ref{fig:Fig_6}, is that we have only $2$ spectral intensity lines at the place of $4$, in the case of the normal spectral function. The normal and anomalous single-particle spectral intensities, presented in Figs.~\ref{fig:Fig_6} and ~\ref{fig:Fig_7}, are of the same order and do not show any sign of the excitonic pair formation in the BLG system. Whether the excitonic condensation is really present at this limit, will be clear after analyzing the two-dimensional map of the anomalous momentum distribution function $g_{b\tilde{a}}({\bf{k}})$ (or the condensate amplitude) in the BLG. The condensate amplitude $g_{b\tilde{a}}({\bf{k}})$, for the zero interlayer interaction limit, is shown in Fig.~\ref{fig:Fig_8}. The momentum distribution function $g_{b\tilde{a}}({\bf{k}})$ shows a prominent ``condensate''-type structure at the corners of unconnected ${\bf{k}}$-cell hexagons in the reciprocal space. These states occur via the interlayer hopping mechanism, as the Eq.(\ref{Equation_37}) suggests. The interlayer hopping amplitude is fixed at $\gamma_{1}/\gamma_{0}=0.128$. In Fig.~\ref{fig:Fig_9}, we have presented the 2D map of the condensate amplitude function $g_{b\tilde{a}}({\bf{k}})$. Thus, the free excitonic pairing states are absent, nevertheless, the function $g_{b\tilde{a}}({\bf{k}})$ shows that the system is in the condensate regime. Therefore, the fundamental excitonic state in the BLG system, at the zero interlayer interaction regime, is governed principally by the electron-hole pair condensed phase caused by the interlayer hopping, and the individual free excitonic pairs are indeed absent in the BLG structure.   
   	
\subsection{\label{sec:Section_4_4} Finite interlayer interactions}
   	
In Figs.~\ref{fig:Fig_10}, \ref{fig:Fig_11} and Fig.~\ref{fig:Fig_12} we have presented the spectral functions ${\cal{S}}_{a}({\bf{k}}, \nu)$, ${\cal{S}}_{b}({\bf{k}}, \nu)$ and  ${\cal{S}}_{b\tilde{a}}({\bf{k}}, \nu)$ for the sufficiently large Coulomb interaction parameter $W=1.25\gamma_0=3.75$ eV. This is the value of $W$, at which the excitonic gap parameter is maximal (see in  Fig.4, in the Ref.\onlinecite{cite_46} ). The interlayer hopping amplitude is fixed at the value $\gamma_1=0.128\gamma_0$ and the zero temperature limit is considered in the pictures. The results, Figs.~\ref{fig:Fig_10}, \ref{fig:Fig_11} and Fig.~\ref{fig:Fig_12}, are of great importance for the whole theory, presented here. We see that there is a large gap in the intensity plots at the Dirac's points $K$ in the BZ of graphene. The double spectral lines corresponding to the electron and hole channels in the normal functions plots, in Figs.~\ref{fig:Fig_10}, \ref{fig:Fig_11}, are well separated now, and the intensities corresponding them are now different. Particularly, the $A$-sublattice electron intensities, (see the spectral function ${\cal{S}}_{a}({\bf{k}}, \nu)$) in Fig.~\ref{fig:Fig_10}, are much higher than the hole intensities in the same picture. Contrary, the electron intensities of the spectral function  ${\cal{S}}_{b}({\bf{k}}, \nu)$, given in Fig.~\ref{fig:Fig_11} are much lower than the hole intensity lines. These results show how the electron-hole pair correlations appear in the bilayer graphene at the finite value of the interlayer Coulomb interaction parameter $W$. Namely, the spectral intensities in Fig.~\ref{fig:Fig_10} show that the sublattice sites $A$ in the lower layer of the BLG are mostly occupied by the electrons, while the $B$-sublattice sites are occupied by the ``hole'' quasiparticles. Thus the electron-hole symmetry in the lower graphene sheet is well pronounced and show that the ``hole'' quasiparticles, at the sites $B$, participate to the formation of the electron-hole pairing states due to the local Coulomb interaction $W$ induced between the quasiparticles at the $B$-sublattice (in the lower layer) and $\tilde{A}$-sublattice (in the upper layer) site positions. The intensity plots in Figs.~\ref{fig:Fig_10}  and ~\ref{fig:Fig_11} are also in well agreement with the intensity plot of the condensate spectral function ${\cal{S}}_{b\tilde{a}}({\bf{k}}, \nu)$, presented in Fig.~\ref{fig:Fig_12}, which shows the probability of the local exciton formation between the quasiparticles at the sublattice sites $B$ and $\tilde{A}$, for a given quantum state $({\bf{k}},\nu)$. The single particle ``hole'' line, in Fig.~\ref{fig:Fig_12}, is much more intense than the electron spectral line, thus corresponds well to the discussion in Figs.~\ref{fig:Fig_10} and ~\ref{fig:Fig_11}.   
   	%
   	\begin{figure}  
   		\begin{center}
   			\includegraphics[width=250px,height=130px]{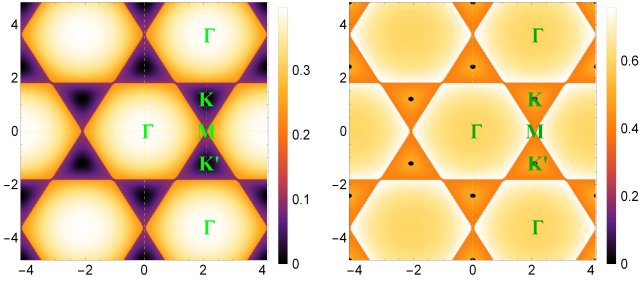}
   			\caption{\label{fig:Fig_13}(Color online) Two-dimensional map of the momentum distribution functions $g_{a}({\bf{k}})$ (see the left panel) and $g_{b}({\bf{k}})$ (see the right panel). The interlayer Coulomb interaction parameter is set at the value $W=1.25\gamma_0$, and the zero temperature limit is considered in the picture. The interlayer hopping amplitude is set at the value $\gamma_1=0.128\gamma_0$.}
   		\end{center}
   	\end{figure} 

   	In Fig.~\ref{fig:Fig_13}, we have shown the $(k_{x},k_{y})$ map of the normal momentum functions, given in Eq.(\ref{Equation_37}). The interlayer interaction parameter is set again at the value $W=1.25\gamma_0$, and the temperature is set at zero. The function $g_{a}({\bf{k}})$ is shown in the left side of the picture, and the function  $g_{b}({\bf{k}})$ is shown in the right panel, in Fig.~\ref{fig:Fig_13}. We see, that the structure of the $(k_{x},k_{y})$ map shows again the possibilities for the excitonic excitations in the system. Particularly, a remarkable particle-hole symmetry is observable in Fig.~\ref{fig:Fig_13}. The bright hexagons on the maps are the regions of the usual single-particle excitations spectra in the system (see, for example, in Ref.\onlinecite{cite_49}, for more details), while the regions between them (see the $6$-apex stars regions) show the particle-hole symmetry in the bottom layer of the BLG. These are the regions, where, potentially, the excitonic condensation could be isolated from the excitonic pairing region (as we will see later on, in this paper).
   	%
   	\begin{figure}  
   		\begin{center}
   			\includegraphics[width=180px,height=160px]{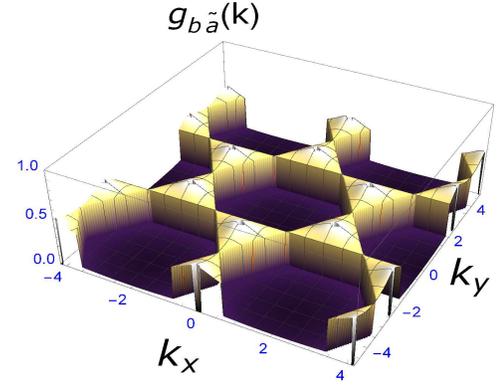}
   			\caption{\label{fig:Fig_14}(Color online) The condensate amplitude function $g_{b\tilde{a}}({\bf{k}})$. The interlayer Coulomb interaction is set at the value $W=1.25\gamma_0$, and the zero temperature limit is considered in the picture. The interlayer hopping amplitude is set at the value $\gamma_1=0.128\gamma_0$.}
   		\end{center}
   	\end{figure} 
   	%
   	\begin{figure}  
   		\begin{center}
   			\includegraphics[width=180px,height=160px]{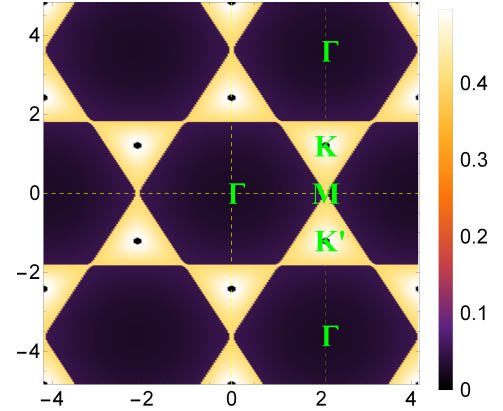}
   			\caption{\label{fig:Fig_15}(Color online) Two-dimensional $(k_{x},k_{y})$ map of the condensate amplitude function $g_{b\tilde{a}}({\bf{k}})$. The interlayer Coulomb interaction parameter is set at the value $W=1.25\gamma_0$, and the zero temperature limit is considered in the picture. The interlayer hopping amplitude is set at the value $\gamma_1=0.128\gamma_0$.}
   		\end{center}
   	\end{figure} 
   	In Figs.~\ref{fig:Fig_14} and ~\ref{fig:Fig_15} we have shown the condensate amplitude function  $g_{b\tilde{a}}({\bf{k}})$ in the bilayer graphene system at the finite interlayer Coulomb interaction parameter $W=1.25\gamma_0$. We see that the excitonic pair formation and condensation principally happens at the values of $k_{x}$ and $k_{y}$, which correspond to the regions, where the particle-hole symmetry is apparent (see the $6$-apex stars regions, in Fig.~\ref{fig:Fig_13}).
   	
   	It is very interesting to see also the behavior of the momentum distribution functions for the very high value of the interlayer interaction parameter $W$. Here, we will consider the value $W=3\gamma_0=9$ eV, (such a value can be realized with the commonly used SiO$_2$ substrate/dielectric with $\epsilon\sim 4$ \cite{cite_8}.)  
   	%
   	\begin{figure}  
   		\begin{center}
   			\includegraphics[width=250px,height=130px]{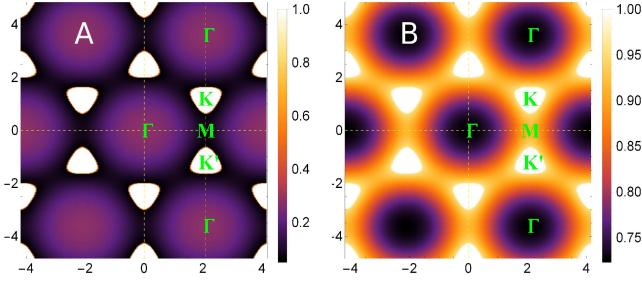}
   			\caption{\label{fig:Fig_16}(Color online) Two-dimensional $(k_{x},k_{y})$ map of the normal momentum distribution functions $g_{a}({\bf{k}})$ (see the left panel) and $g_{b}({\bf{k}})$ (see the right panel). The interlayer Coulomb interaction parameter is set at the value $W=3\gamma_0$, and the zero temperature limit is considered in the picture. The interlayer hopping amplitude is set at the value $\gamma_1=0.128\gamma_0$.}
   		\end{center}
   	\end{figure} 
In Fig.~\ref{fig:Fig_16} we have plotted the normal momentum distribution functions for $W=3\gamma_0$. The left panel corresponds to the function $g_{a}({\bf{k}})$ and the right panel shows the ${\bf{k}}$-dependence of the function $g_{b}({\bf{k}})$.
First of all, the $6$-apex stars regions, in the normal momentum distribution spectra, do not reflect anymore the particle-hole symmetry in this case. Thus, deep in these regions, the excitonic pairing could not happen. Contrary, now, the whole quasiparticle excitation regions (see the ${\bf{k}}$-cell hexagonal regions, in Fig.~\ref{fig:Fig_16}, marked as $A$ and $B$) represent the possible regions of the wave vector components $k_x$ and $k_y$, at which the excitonic pair formation could appear, and these are the regions now, which represent the particle-hole symmetry in the problem. We see, in Fig.~\ref{fig:Fig_17}, that the excitonic pair formation regions (the corresponding momentum functions have very small amplitudes, as compared to the weak interaction regime), in the BLG, cover nearly the entire ${\bf{k}}$-space, except the regions, corresponding to the $6$-apex stars pockets. In addition, we see, in Fig.~\ref{fig:Fig_17}, how the excitonic condensation peaks surround the free excitonic pair formation hexagonal ${\bf{k}}$-cell regions, just at the borders of the dark triangular Dirac's pockets. The 2D map of the function $g_{b\tilde{a}}({\bf{k}})$ at $W=3\gamma_0$ is shown in Fig.~\ref{fig:Fig_18}). 
   	%
   	\begin{figure}  
   		\begin{center}
   			\includegraphics[width=180px,height=160px]{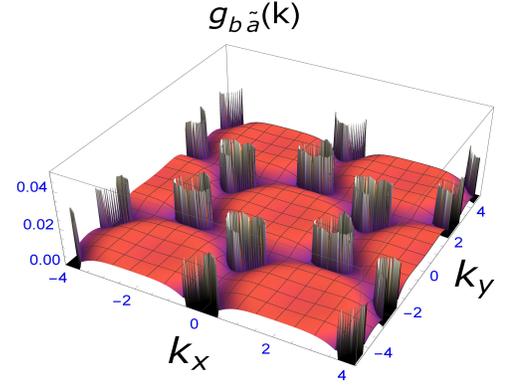}
   			\caption{\label{fig:Fig_17}(Color online) The condensate amplitude function $g_{b\tilde{a}}({\bf{k}})$ at $T=0$. The interlayer Coulomb interaction is set at the value $W=3\gamma_0$ and the interlayer hopping amplitude is $\gamma_1=0.128\gamma_0$.}
   		\end{center}
   	\end{figure} 
   	%
   	%
   	\begin{figure}  
   		\begin{center}
   			\includegraphics[width=200px,height=160px]{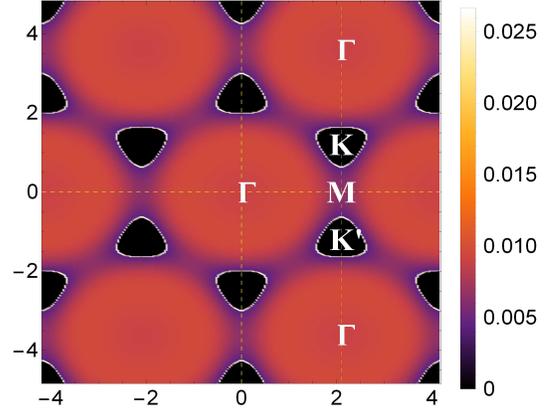}
   			\caption{\label{fig:Fig_18}(Color online) Two-dimensional $(k_{x},k_{y})$ map of the condensate amplitude function $g_{b\tilde{a}}({\bf{k}})$. The interlayer Coulomb interaction parameter is set at the value $W=3\gamma_0$, and the zero temperature limit is considered in the picture. The interlayer hopping amplitude is set at the value $\gamma_1=0.128\gamma_0$.}
   		\end{center}
   	\end{figure} 
   	%
   	\subsection{\label{sec:Section_4_5} The ${\bf{k}}$-space evolution of the condensate states}
   	%
The principal conclusion that could be gained from the results in Figs.~\ref{fig:Fig_17} and ~\ref{fig:Fig_18} is the following: at the high values of the interlayer Coulomb interaction parameter $W$, we have the separation of the excitonic condensate states from the free excitonic pair formation regions in the interacting BLG, and the system is apparently in the strong condensate regime, which coexists with the excitonic insulator state. Thus, at large $W$, we have a semiconductor with the well defined interband and intraband band gaps of the hole-decay spectrum, and the BLG is characterized by the mixed states, composed of the excitonic pair formation and condensate states. Another fundamental observation from the obtained numerical results is related to the existence of the excitonic BCS-Bose-Einstein-Condensation (BEC)-like crossover mechanism in the BLG system, where the tunable parameter is the interlayer Coulomb interaction $W$. The corresponding experimental parameter, that could tune a similar transition is the additional charge density on both layers of BLG (see the works \cite{cite_38, cite_39, cite_40}).
   	
\begin{figure*}  
   	 		\begin{center}
   	 			\includegraphics[width=500px,height=120px]{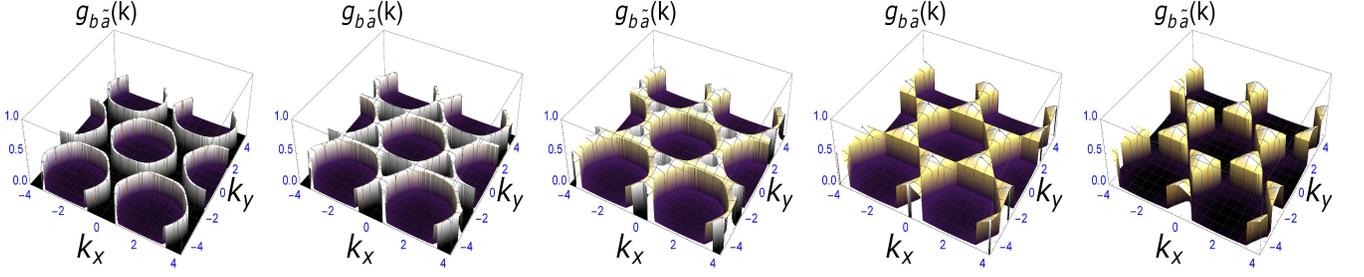}
   	 			\caption{\label{fig:Fig_19}(Color online) The condensate amplitude function $g_{b\tilde{a}}({\bf{k}})$ for different values of the interlayer Coulomb interaction parameter $W$ ($W=0.5\gamma_0$, $W=0.8\gamma_0$, $W=1.1\gamma_0$, $W=1.25\gamma_0$, $W=1.3\gamma_0$, from left to right). The zero temperature limit is considered in the picture. The interlayer hopping amplitude is set at the value $\gamma_1=0.128\gamma_0$.}
   	 		\end{center}
   	 	\end{figure*}

\begin{figure*}  
   	  		\begin{center}
   	  			\includegraphics[width=500px,height=120px]{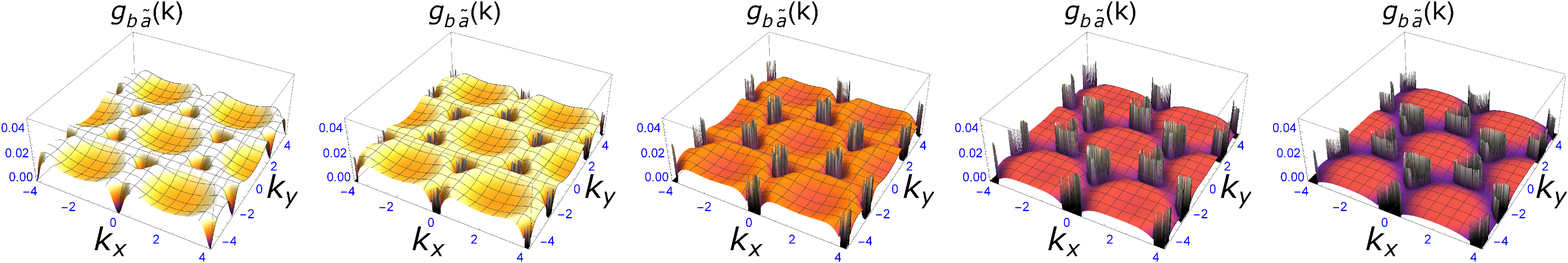}
   	  			\caption{\label{fig:Fig_20}(Color online) The condensate amplitude function $g_{b\tilde{a}}({\bf{k}})$ for different values of the interlayer Coulomb interaction parameter $W$ ($W=1.5\gamma_0$, $W=1.7\gamma_0$, $W=2\gamma_0$, $W=3\gamma_0$, $W=4\gamma_0$, from left to right). The zero temperature limit is considered in the picture. The interlayer hopping amplitude is set at the value $\gamma_1=0.128\gamma_0$.}
   	  		\end{center}
   	  	\end{figure*}

Here, we would like to demonstrate also, the ${\bf{k}}$-space evolution of the condensate states appearing (in the semiconducting phase) around the Dirac's points $K$ in the reciprocal space by forming the condensate pockets (Dirac's pockets) in the $\bf{k}$-space. In Figs.~\ref{fig:Fig_19} and ~\ref{fig:Fig_20}, we have shown the evolution of the anomalous momentum distribution function as a function of the interlayer Coulomb interaction parameter $W$. A broad interval of values of the interaction parameter $W$ has been considered: from very weak interaction limit: $W=0.5\gamma_0$(see in the first panel, in Fig.~\ref{fig:Fig_19}), up to very high value of $W$: $W=4\gamma_0$ (see in the last panel, in Fig.~\ref{fig:Fig_20}).

   	 	%
   	  	%
   	  	 	\begin{figure*}  
   	  	 		\begin{center}
   	  	 			\includegraphics[width=500px,height=110px]{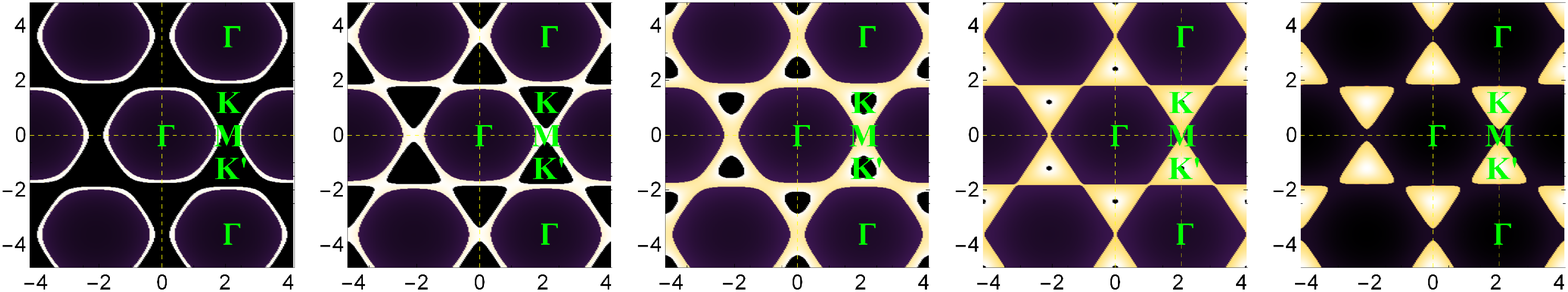}
   	  	 			\caption{\label{fig:Fig_21}(Color online) The 2D ${\bf{k}}$-map of the condensate amplitude function $g_{b\tilde{a}}({\bf{k}})$ for different values of the interlayer interaction parameter $W$ ($W=0.5\gamma_0$, $W=0.8\gamma_0$, $W=1.1\gamma_0$, $W=1.25\gamma_0$, $W=1.3\gamma_0$, from left to right). The zero temperature limit is considered in the picture. The interlayer hopping amplitude is set at the value $\gamma_1=0.128\gamma_0$.}
   	  	 		\end{center}
   	  	 	\end{figure*} 
   	  	 	%
   	  	 	\begin{figure*}  
   	  	 		\begin{center}
   	  	 			\includegraphics[width=500px,height=110px]{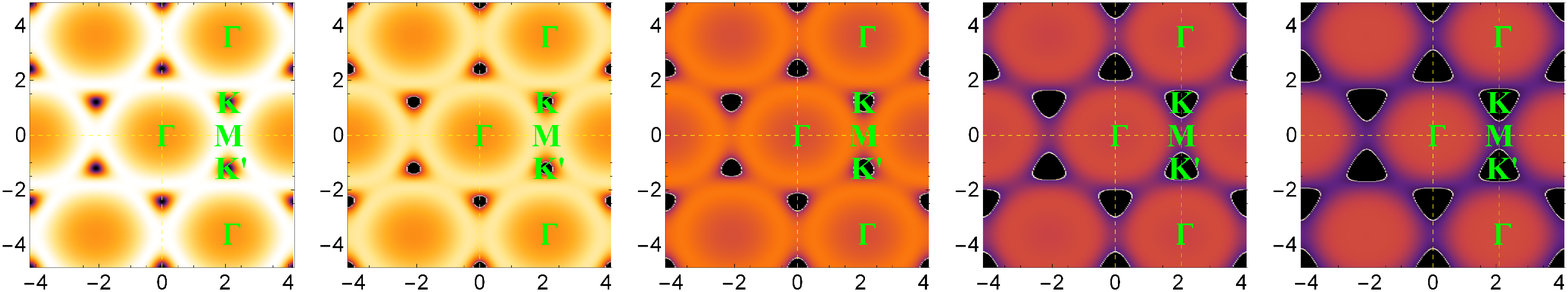}
   	  	 			\caption{\label{fig:Fig_22}(Color online) The 2D ${\bf{k}}$-map of the condensate amplitude function $g_{b\tilde{a}}({\bf{k}})$ for different values of the interlayer interaction parameter $W$ ($W=1.5\gamma_0$, $W=1.7\gamma_0$, $W=2\gamma_0$, $W=3\gamma_0$, $W=4\gamma_0$, from left to right). The zero temperature limit is considered in the picture. The interlayer hopping amplitude is set at the value $\gamma_1=0.128\gamma_0$.}
   	  	 		\end{center}
   	  	 	\end{figure*}

As it was discussed and explained previously, at the zero interaction limit, the BLG system is in the excitonic pair condensate regime, mediated by the local interlayer hopping parameter $\gamma_1$. For the finite interlayer interaction value $W=0.132\gamma_0=0.396$ eV, a finite hybridization gap opens in the bilayer graphene. We have shown that for a sufficiently large interval of $W$ (when $W\in(0.0,1.3\gamma_0)$) the BCS-like weak coupling theory is valid for the BLG. In this case, we converge to the low-energy results, given in Refs.\onlinecite{cite_10, cite_11, cite_12, cite_13,cite_14}, where the weak coupling BCS theory is evaluated to obtain the excitonic pairing gap parameter. This has the similarities also with the 2D square lattice excitonic systems \cite{cite_24, cite_25, cite_26, cite_27}, where the excitonic insulator state, at the low interband Coulomb interaction limit, is governed by the BCS-like pair condensed state, which is transformed into the excitonic BEC state at the large interband $U$-interaction limit. When augmenting the interlayer Coulomb interaction parameter (see already in the second panel, in Fig.~\ref{fig:Fig_19}, where $W/\gamma_0=0.8$), the ${\bf{k}}$-cell hexagons are merging at the  $M$ symmetry points in the reciprocal space (see also in the second panel, in Fig.~\ref{fig:Fig_21}, for the 2D map of the anomalous momentum distribution function), and the system passes into the mixed state, composed of the excitonic pairs. Particularly, the Dirac's pockets start to develop, which surround the Dirac's neutrality points $K$ and $K'$. Furthermore, when increasing the parameter $W$ (see in the rest of the panels in Fig.~\ref{fig:Fig_19}, where $W=1.\gamma_0, W=1.25\gamma_0, W=1.3\gamma_0$), the excitonic pair formation states are replenishing the ${\bf{k}}$-space pockets. At the value $W=1.3\gamma_0$, (see in the last panel, in Fig.~\ref{fig:Fig_19}, and also in Fig.~\ref{fig:Fig_21}) the Dirac's pockets in the ${\bf{k}}$-space are completely replenished, and are isolated from each other, along the directions $K-M-K'$, in the reciprocal space. At this value of $W$, the bilayer graphene is in the semiconducting phase, as the electronic band structure shows (see in the Section \ref{sec:Section_3_2} above). At the values of $W$, corresponding to the interval $W>W_j$, where $W_j$ is the value at which the chemical potential passes into its upper bound (with a sufficiently large jump), the excitonic condensates states appear and start to separate from the excitonic pair formation regions (see the concave like structures,  which appear in the middle of the Dirac's pockets, in the first three panels in Fig.~\ref{fig:Fig_20} and also in Fig.~\ref{fig:Fig_22}) at the cost of breaking of the mixed states. When continuing to increase the interlayer Coulomb interaction parameter, the excitonic condensation peaks appear in the reciprocal ${\bf{k}}$-space at the borders of the pockets regions, and the mixed state continues to break. Simultaneously, we see that the free excitonic pair formation regions are formed (see the concave like bright hexagonal regions in Figs.~\ref{fig:Fig_20} and ~\ref{fig:Fig_22}) at the place of the empty dark hexagonal regions in ${\bf{k}}$-space (see in the panels in Figs.~\ref{fig:Fig_19} and ~\ref{fig:Fig_21}).
   	
In the last panels, in Figs.~\ref{fig:Fig_20} and ~\ref{fig:Fig_22}, we have shown the condensate amplitude function for the very high value of the interaction parameter: $W=4\gamma_0$. We observe in Fig.~\ref{fig:Fig_20} that the amplitude of the anomalous momentum function is considerably reduced, and the concave-like excitonic pair formation regions become convex-like, improving the free excitonic pairing state in the system (like in the semiconducting systems). In addition, there is a strong evidence for the excitonic condensation in the BLG, and the excitonic condensates states are completely separated from the excitonic pair formation regions (see the condensates peaks at the borders of the Dirac's pockets, in the reciprocal space, in Fig.~\ref{fig:Fig_20}). Thus, at the high values of the interlayer Coulomb interaction, the excitonic condensate states are of the BEC-type, and we observe a remarkable BCS-BEC-like crossover mechanism in the bilayer graphene system, analogue to that observed in some intermediate valent semiconductor systems \cite{cite_24, cite_25, cite_26, cite_27}, where the similar crossover is due to the interband Coulomb interaction parameter $U$. Note, that the BCS-BEC type crossover in the extrinsic BLG structure is observed also recently in Ref.\onlinecite{cite_8}, in the presence of the perpendicular magnetic field, when the localized magnetoexcitons form.
   	
We see particularly, in Fig.~\ref{fig:Fig_22}, how the excitonic condensate formation develops in the reciprocal space, at the values $W>W_j$ by forming the Dirac's pockets around the symmetry points $K$ and $K'$. In the panels in Figs.~\ref{fig:Fig_21}  and ~\ref{fig:Fig_22} we see clearly how the dark hexagonal regions in Fig.~\ref{fig:Fig_21}, are transforming into the bright hexagonal ones and the excitonic pairing states with small amplitudes are formed, and, at the same time, the condensate picks appear in the momentum space, which surround the holy Dirac's pockets in the ${\bf{k}}$-space (see, for example, in the last panel, in Fig.~\ref{fig:Fig_22}).   
%
\section{\label{sec:Section_5} Conclusion}
%
We have studied the influence of the excitonic effects on the spectral properties of the bilayer graphene. 
The theory, evaluated in the present paper, permits to construct a fully controllable theoretical model for the studies of the excitonic condensation effects in the bilayer graphene system and to manipulate a number of physical parameters, which are important from theoretical and experimental points of views. 
   	
We have reconstructed the electronic band structure of the bilayer graphene, taking into account the local excitonic pairing interaction between the layers of the system. We have shown that a remarkable hybridization gap appears in the electronic spectrum at sufficiently small values of the interlayer coupling parameter. We estimated the values of the hybridization gap for a large interval of the interaction parameter. We have demonstrated that there is a semimetal-semiconductor transition in the BLG system, tuned by the local Coulomb interaction parameter. We have shown that the chemical potential solution has two bounds: lower and upper, which controls the borders of the excitonic pair formation and excitonic condensate phases. Meanwhile, the additional effective chemical potential $\bar{\mu}$ appears in the energy spectrum of the bilayer graphene, which plays the role of the true Fermi energy in the bilayer graphene system. This result explains, at least qualitatively, the recent experimental effort to determine the chemical potential behavior in the gated bilayer graphene systems \cite{cite_38, cite_39, cite_40}. The particle-hole symmetry, in the single layer of bilayer graphene, becomes strongly apparent when analyzing the normal and anomalous momentum distribution functions properties. Namely, in the regions, where the particle-hole symmetry is conserved in the normal distribution function, the excitonic pairing occurs at the non-zero values of the local interlayer Coulomb interaction parameter. In this limit of the very small and intermediate values of the interlayer Coulomb interaction, the double layer system is governed by the effective weak coupling BCS-like pairing states \cite{cite_50, cite_51}, and the excitonic coherence length satisfies well the BCS-like relation $\xi_c=\alpha\bar{\mu}/({k_{\cal{F}}}\Delta)$, with $\Delta$, being the excitonic pairing gap parameter. Contrary, in strong coupling limit the coherence length becomes proportional to $\Delta$ ($\xi_c\sim\Delta)$, and the system is in the excitonic condensate regime in that case. By analysing the anomalous momentum distribution functions properties (directly related to the excitonic pair formation or condensation) and by varying the interlayer interaction parameter (from very small up to very high values), we have found a remarkable BCS-BEC type crossover, which is different (at its origin) from the analogue transitions obtained in the intermediate-valent semiconductors, or rare-earth compounds \cite{cite_19, cite_20, cite_21,cite_22,cite_23,cite_24,cite_25,cite_26}. We have shown the evolution of the excitonic pairing and condensate states as a function of the Coulomb interaction parameter and we have found the regions, where the system is in the excitonic pairing or condensate regimes. It is remarkable to mention that, for the large values of the coupling parameter, the excitonic condensate states appears around the Dirac's points $K$ and $K'$, by creating the Dirac's ``hole'' pockets in the reciprocal space. Namely, triangular ${\bf{k}}$-pockets are formed, around which the new excitonic BEC states appear and coexist with the excitonic pairing states, which cover the large regions in the ${\bf{k}}$-space (creating by this the mixed region: excitonic pairing+excitonic BEC). It is remarkable to note also that the surface forms, corresponding to the free excitonic pair formation regions (the excitonic insulator state), transform from the concave-like into the convex-like when augmenting the interlayer Coulomb interaction parameter.
   	
The results obtained in the present paper show that the spectral functions and the momentum distribution functions provide a direct proof of the excitonic effects in the bilayer graphene system. On the other hand, the experimental measurement of the anomalous functions is extremely difficult, due to the very short lifetime of excitonic quasiparticles and fast electron-hole recombination effects. We hope that these difficulties, related to the measurements of the anomalous momentum distribution functions, will be overcome in the near future by the improvement of the bilayer graphene fabricating techniques and by the setup architecture of the biased double-layer heterostructures. The principal advantage for such a success would be the simultaneous measurements of the photon's absorption and emission spectra in different layers of the BLG. The results of the presented paper are particularly important in the context of the recent experimental results \cite{cite_52, cite_53} concerning the discovery of the laser-induced white light emission spectra in graphene ceramics. It was demonstrated in Ref.\cite{cite_52} that a large bandgap opening and light emission from graphene is possible by using the continuous-wave laser beams with wavelengths from the visible to the near-infrared range. This hidden multistability of graphene is fundamental to create a semiconducting phase immersed in the semimetallic continuum one. 
\appendix
   	

%

\begin{thebibliography}{10}
%
\bibitem{cite_1} W. Ehrenberg, Electric conduction in semiconductors and metals, Oxford University Press, 1958.
\bibitem{cite_2} A.H. Castro Neto, F. Guinea, N.M.R. Peres, K.S. Novoselov, A.K. Geim, Rev. Mod. Phys. 81 (2009) 109.
\bibitem{cite_3} Eduardo V. Castro, K. S. Novoselov, S. V. Morozov, N.M.R. Peres, J.M.B. Lopes dos Santos, Johan Nilsson, F. Guinea, A.K. Geim, A.H. Castro Neto, Phys. Rev. Lett. 99 (2007) 216802.
\bibitem{cite_4} J.P. Eisenstein, A.H. MacDonald, Nature 432 (2004) 691.
\bibitem{cite_5} J.J. Su, A.H. MacDonald, Nat. Phys. 4 (2008) 799.
\bibitem{cite_6} D.S.L. Abergel,  V. Apalkov, J. Berashevich, K. Ziegler, Tapash Chakraborty, Adv. Phys. 59 (2010) 261.
\bibitem{cite_7} R. Saito, G. Dresselhaus, M.S. Dresselhaus, Physical Properties of Carbon Nanotubes, Imperial
   		College Press, London 1998.
\bibitem{cite_8} Van-Nham Phan, H. Fehske, New Journal of Physics 14 (2012) 075007.
\bibitem{cite_9} C.H. Zhang, Y.N. Joglekar, Phys. Rev. B 77 (2008) 233405.
\bibitem{cite_10} H. Min, R. Bistritzer, J.J. Su, A.H. MacDonald, Phys. Rev. B 78 (2008) 121401.
\bibitem{cite_11} Y.E. Lozovik, A.A. Sokolik, JETP Lett. 87 (2008) 55.
\bibitem{cite_12} Yu.E. Lozovik, S.L. Ogarkov, A.A. Sokolik, Phys. Rev. B 86 (2012) 045429.
\bibitem{cite_13} M.Yu. Kharitonov, K.B. Efetov, Phys. Rev. B 78 (2008) 241401(R).
\bibitem{cite_14} M.Yu. Kharitonov, K.B. Efetov, Semicond. Sci. Technol. 25 (2010) 034004.
\bibitem{cite_15} J. Neuenschwander, P. Wachter, Phys. Rev. B 41 (1990) 12693.
\bibitem{cite_16} B. Bucher, P. Steiner, P. Wachter, Phys. Rev. Lett. 67 (1991) 2717 .
\bibitem{cite_17} P. Wachter, Solid State Commun. 118 (2001) 645.
\bibitem{cite_18} P. Wachter, B. Bucher, J. Malar, Phys. Rev. B 69 (2004) 094502.
\bibitem{cite_19} L.V. Keldysh, Y.V. Kopaev, Fiz. Tverd. Tela (Leningrad) 6 (1964) 2791  [Sov. Phys. Solid State 6 (1965) 2219].
\bibitem{cite_20} J. des Cloizeaux, J. Chem. Phys. Solids. 26 (1965) 259.
\bibitem{cite_21} W. Kohn, in Many Body Physics, edited by C. de Witt, R. Balian, Gordon Breach, New York, 1968.
\bibitem{cite_22} D. J\'{e}rome, T.M. Rice, W. Kohn, Phys. Rev. 158 (1967) 462.
\bibitem{cite_23} A. Griffin, D.W. Snoke, S. Stringari, eds Bose-Einstein Condensation, Cambridge University Press, 1995.
\bibitem{cite_24} K. Seki, R. Eder, Y. Ohta, Phys. Rev. B 84 (2011) 245106.
\bibitem{cite_25} B. Zenker, D. Ihle, F.X. Bronold, H. Fehske, Phys. Rev. B 85 (2012) 121102(R).
\bibitem{cite_26} B. Zenker, D. Ihle, F.X. Bronold, H. Fehske, Phys. Rev. B 81 (2010) 115122.
\bibitem{cite_27} B. Zenker, D. Ihle, F.X. Bronold, H. Fehske, Phys. Rev. B 83 (2011) 235123.
\bibitem{cite_28} B. Rosenstein, M. Lewkowicz, T. Maniv, Phys. Rev. Lett. 110 (2013) 066602.
\bibitem{cite_29}  J.E. Drut, T.A. Lahde, Phys. Rev. Lett. 102 (2009) 026802; Phys. Rev. B 79 (2009) 165425 .
\bibitem{cite_30} G.W. Semenoff, Phys. Rev. Lett. 53 (1984) 2449.
\bibitem{cite_31} H.K. Min, B. Sahu, S.K. Banerjee, A.H. MacDonald, Phys. Rev. B 75 (2007) 155115.
\bibitem{cite_32} F. Wang, Y. Zhang, C. Tian, C. Girit, A. Zettl, M. Crommie, Y.R. Shen, Science 320, (2008) 206.
\bibitem{cite_33} Z.Q. Li, E.A. Henriksen, Z. Jiang, Z. Hao, M.C. Martin, P. Kim, H.L. Stormer, D.N. Basov, Phys. Rev. Lett. 102 (2009) 037403.
\bibitem{cite_34} Y. Zhang, Tsung-Ta Tang, C. Girit, Zhao Hao, M.C. Martin, A. Zettl, M.F. Crommie, Y. Ron Shen, F. Wang, Nature 459, (2009) 820.
\bibitem{cite_35} M.S. Dresselhaus, G. Dresselhaus, Adv. Phys. 51 (2002) 1.
\bibitem{cite_36} J.W. Negele, H. Orland, Quantum Many-Particle Systems, Addison-Wesley, Reading, MA, 1988.
\bibitem{cite_37} A.A. Abrikosov, L.P. Gorkov, I.E. Dzyaloshinski, Methods of Quantum Field Theory in Statistical Physics, Pergamon Press, 1965.
\bibitem{cite_38} K. Lee, B. Fallahazad, J. Xue, D.C. Dillen, K. Kim, T. Taniguchi, K. Watanabe, E. Tutuc, Science 345 (2014) 58.
\bibitem{cite_39} S. Kim, I. Jo, J. Nah, Z. Yao, S.K. Banerjee, E. Tutuc, Phys. Rev. B 83 (2011) 161401(R).
\bibitem{cite_40} S. Kim, I. Jo, D.C. Dillen, D.A. Ferrer, B. Fallahazad, Z. Yao, S.K. Banerjee, E. Tutuc, Phys. Rev. Lett. 108 (2012) 116404.
\bibitem{cite_41} Z.F. Wang, Q. Li, H. Su, X. Wang, Q.W. Shi, J. Chen, J. Yang, J.G. Hou, Phys.
Rev. B, 75, (2007) 085424.
\bibitem{cite_42} M. Mucha-Kruczynski, E. McCann, V.I. Fal'ko, Semicond. Sci. Technol. 25 (2010) 033001.
\bibitem{cite_43} C. Bendtsen, O. Stauning, FADBAD, a Flexible C++ Package for Automatic Differentiation,
Department of Mathematical Modelling, Technical University of Denmark, 1996.	
\bibitem{cite_44} E. McCann, M. Koshino, Rep. Prog. Phys. 76 (2013) 056503.
\bibitem{cite_45} T. Ohta, A. Bostwick, T. Seyller, K. Horn, E. Rotenberg, Science 951 (2006) 4.
\bibitem{cite_46} V. Apinyan, T.K. Kope\'{c}, Phys. Scr. 91 (2016) 095801.
\bibitem{cite_47} S. Kim, I. Jo, J. Nah, Z. Yao, S.K. Banerjee, E. Tutuc,  Phys. Rev. B 83 (2011) 161401(R).
\bibitem{cite_48} H. Min,  B. Sahu, S.K Banerjee, A.H. MacDonald, Phys. Rev. B 75 (2007) 155115.
\bibitem{cite_49} A. Bostwick, T. Ohta, Th. Seyller, K. Horn, E. Rotenberg, Science 313 (2006) 951.
\bibitem{cite_50} D.S.L. Abergel, M. Rodriguez-Vega, E. Rossi, S. Das Sarma, Phys. Rev. B 88 (2013) 235402.
\bibitem{cite_51} D.S.L. Abergel, R. Sensarma, S. Das Sarma, Phys. Rev. B 86 (2012) 155447(R).
\bibitem{cite_52} W. Strek, R. Tomala, M. Lukaszewicz, B. Cichy, Y. Gerasymchuk, P. Gluchowski, L. Marciniak, A. Bednarkiewicz, D. Hreniak, Scientific Reports 7 (2017) 41281.
\bibitem{cite_53} W. Strek, B. Cichy, L. Radosinski, P. Gluchowski, L. Marciniak, M. Lukaszewicz, Dariusz Hreniak, Light: Science and Applications 4 (2015) 237.
   			
  		
\end{thebibliography}
\end{document}